\documentclass[twoside,11pt,final]{article}  % Specifies the document style.
\usepackage{epsfig}
\usepackage{amssymb}

\newcommand{\ftn}{\footnotesize}

\newcommand{\ssz}{\scriptsize}

\newcommand{\tr}{{\mbox{\sf\ssz T}}}
\newcommand{\Tr}{\mbox{\sf Tr}}

\newcommand{\hepph}[1]{{\tt hep-ph/#1}}
\newcommand{\hepex}[1]{{\tt hep-ex/#1}}
\newcommand{\astroph}[1]{{\tt astro-ph/#1}}

\newcommand{\Fia}{\mbox{
$\matrix{\llgm{d_r&-u_r \cr d_r&-u_r \cr d_r&-u_r\cr
e_r&-\nu_r}\rrgm}$}}

\newcommand{\Fca}{\mbox{
$\matrix{\llgm{u^c_r&u^c_r&u^c_r&\nu^c_r\cr
d^c_r&d^c_r&d^c_r&e^c_r}\rrgm}$}}

\newcommand{\Hcb}{\mbox{
$\matrix{\llgm{\bar u^c_H&\bar d^c_H \cr \bar u^c_H&\bar d^c_H \cr
\bar u^c_H&\bar d^c_H \cr \bar \nu^c_H&\bar e^c_H }\rrgm}$}}

\newcommand{\Hca}{\mbox{
$\matrix{\llgm{u^c_H&u^c_H&u^c_H&\nu^c_H\cr
d^c_H&d^c_H&d^c_H&e^c_H}\rrgm}$}}

\newcommand{\bdh}{\mbox{
$\llgm{h^+_2 & h^0_1 \cr h^0_2 & h^-_1 }\rrgm$}}

\def\to{\rightarrow}

\def\llgm{\left\lgroup\matrix}
\def\rrgm{\right\rgroup}

\def\beq{\begin{equation}}
\def\eeq{\end{equation}}
\def\bea{\begin{eqnarray}}
\def\eea{\end{eqnarray}}
\def\openep{\leavevmode\hbox{\normalsize$\epsilon$\kern-4.pt$\epsilon$}}

\begin{document}

\title{\sc Supersymmetric Dark Matter, \\ Inflation
and Yukawa Quasi-Unification}
\author{{\sc G. Lazarides and C. Pallis}\\ \\
{\sl\ftn Physics Division, School of Technology,}\\
{\sl\ftn Aristotle University of Thessaloniki,} \\
{\sl\ftn 54124 Thessaloniki, GREECE} \\
{\tt\ftn lazaride@eng.auth.gr, kpallis@auth.gr}}

\maketitle

\pagestyle{myheadings}

\thispagestyle{plain}\markboth{G. Lazarides and C.
Pallis}{Supersymmetric Dark Matter, Inflation and Yukawa
Quasi-Unification} \setcounter{page}{1}

\begin{abstract}

{\ftn The construction of specific supersymmetric grand unified
models based on the Pati-Salam gauge group and leading to a set of
Yukawa quasi-unification conditions is briefly reviewed. For each
sign of the parameter $\mu$, an appropriately chosen condition
from this set can allow an acceptable $b$-quark mass within the
constrained minimal supersymmetric standard model. The
restrictions on the parameter space which arise from the cold dark
matter constraint, the inclusive decay $b\rightarrow s\gamma$, the
muon anomalous magnetic moment and the collider bounds are also
investigated. For $\mu>0$, a wide and natural range of parameters
is allowed. On the contrary, the $\mu<0$ case not only is
disfavored by the present experimental data on the muon anomalous
magnetic moment, but also is excluded by the combination of the
cold dark matter and $b\to s \gamma$ requirements. In the $\mu>0$
case, the predicted neutralinos are possibly detectable in the
future direct cold dark matter searches. Moreover, the $\mu$ term
is generated via a Peccei-Quinn symmetry and proton is practically
stable. The same model gives rise to a new shifted inflationary
scenario, which is based only on renormalizable terms, does not
suffer from the problem of monopole overproduction at the end of
inflation and is compatible with the cosmic microwave background
radiation constraint. Although the relevant part of inflation
takes place at values of the inflaton field which are not much
smaller than the reduced Planck scale and, thus, supergravity
corrections could easily invalidate inflation, this does not
happen provided that an extra gauge singlet with a superheavy
vacuum expectation value, which originates from D-terms, is
introduced and a specific form of the K\"{a}hler potential is
used. The constraint from the cosmic background explorer can again
be met by readjusting the values of the parameters which were
obtained with global supersymmetry. }

\end{abstract}

\newpage

\section{Introduction}\label{sec:intro}

The constrained minimal supersymmetric standard model (CMSSM),
which is a highly predictive version of the minimal supersymmetric
standard model (MSSM) based on universal boundary conditions
\cite{Cmssm}, can be further restricted by being embedded in a
supersymmetric (SUSY) grand unified theory (GUT) with a gauge
group containing $SU(4)_{\rm c}$ and $SU(2)_R$. This can lead
\cite{pana} to `asymptotic' Yukawa unification (YU) \cite{als},
i.e. the exact unification of the third generation Yukawa coupling
constants at the SUSY GUT scale $M_{\rm GUT}$. Indeed, assuming
that the electroweak Higgs superfields $H_1$, $H_2$ and the third
family right handed quark superfields $t^c$, $b^c$ form $SU(2)_R$
doublets, we obtain \cite{pana} the asymptotic Yukawa coupling
relation $h_t=h_b$ and, hence, large $\tan\beta\sim m_{t}/m_{b}$.
Moreover, if the third generation quark and lepton $SU(2)_L$
doublets [singlets] $q_3$ and $l_3$ [$b^c$ and $\tau^c$] form a
$SU(4)_{\rm c}$ {\bf 4}-plet [${\bf\bar 4}$-plet] and the Higgs
doublet $H_1$ which couples to them is a $SU(4)_{\rm c}$ singlet,
we get $h_b=h_{\tau}$ and the asymptotic relation $m_{b}=m_{\tau}$
follows. The simplest GUT gauge group which contains both
$SU(4)_{\rm c}$ and $SU(2)_R$ is the Pati-Salam (PS) group $G_{\rm
PS}=SU(4)_c\times SU(2)_L\times SU(2)_R$ (for YU within $SO(10)$,
see Ref. \cite{rabyr}).

However, given the experimental values of the top-quark and
tau-lepton masses (which, combined with YU, naturally restrict
$\tan\beta\sim50$), the CMSSM supplemented by the assumption of YU
yields unacceptable values of the $b$-quark mass for both signs of
the parameter $\mu$. This is due to the generation of sizeable
SUSY corrections \cite{copw} to $m_b$ (about 20$\%$), which arise
from sbottom-gluino (mainly) and stop-chargino loops \cite{copw,
pierce, susy} and have the sign of $\mu$ (with the standard sign
convention of Ref. \cite{sugra}). The predicted tree-level
$m_b(M_Z)$, which turns out to be close to the upper edge of its
$95\%$ confidence level (c.l.) experimental range
\beq 2.684~{\rm GeV}\lesssim  m_b(M_Z) \lesssim 3.092~{\rm
GeV}~~(\mbox{with}~~\alpha_s(M_Z)=0.1185),\label{mbrg}\eeq
receives, for $\mu>0$ [$\mu<0$], large positive [negative]
corrections which drive it well above [a little below] the allowed
range. This range is derived \cite{qcdm} from the $95\%$ c.l.
range for $m_b(m_b)$ in the $\overline{MS}$ scheme
$(3.95-4.55~{\rm GeV})$ \cite{mb} evolved up to $M_Z$ in accord
with the analysis of Ref. \cite{baermb}. Consequently, for both
signs of $\mu$, YU leads to an unacceptable $m_b(M_Z)$, with the
$\mu<0$ case being much less disfavored.

The usual strategy to resolve this discrepancy is the introduction
of several kinds of nonuniversalities in the scalar \cite{king,
raby, baery} and/or gaugino \cite{nath,su5b} sector of MSSM with
an approximate preservation of YU. On the contrary, in Ref.
\cite{qcdm}, this problem is addressed in the context of the PS
GUT, without need of invoking departure from the CMSSM
universality. The Higgs sector of the model is extended by
including an extra $SU(4)_{\rm c}$ non-singlet Higgs superfield
with Yukawa couplings to the quarks and leptons. The Higgs
$SU(2)_L$ doublets contained in this superfield can naturally
develop \cite{wetterich} subdominant vacuum expectation values
(VEVs) and mix with the main electroweak doublets which are
assumed to be $SU(4)_{\rm c}$ singlets and form a $SU(2)_R$
doublet. This mixing can, in general, violate $SU(2)_R$.
Consequently, the resulting electroweak doublets $H_1$, $H_2$ do
not form a $SU(2)_R$ doublet and break $SU(4)_{\rm c}$ too. As a
consequence, a moderate violation of the YU is obtained, which can
allow an acceptable $b$-quark mass even with universal boundary
conditions. Obviously, a small deviation from YU is enough for an
acceptable prediction of the $b$-quark mass when $\mu<0$, while,
for $\mu>0$, a more pronounced deviation is needed. \vskip0.10cm

In this review, we outline the construction of two particular SUSY
GUT models which can cause a deviation from YU adequate for
$\mu>0~[\mu<0]$ in the first [second] model. We then discuss the
resulting CMSSM in each case and the various astrophysical and
experimental requirements which restrict its parameter space. They
originate from the data on the cold dark matter (CDM) abundance in
the universe derived by the Wilkinson microwave anisotropy probe
(WMAP) \cite{wmap, wmapl}, the inclusive branching ratio of
$b\rightarrow s\gamma$ \cite{cleo}, ${\rm BR}(b\to s\gamma)$, the
muon anomalous magnetic moment $\alpha_\mu$ \cite{muon} and the
mass of the lightest Higgs boson $m_h$ \cite{higgs}. We show that,
for $\mu>0$, our model possesses a wide range of parameters which
is consistent with all these constraints. On the contrary, for
$\mu<0$, the upper bound on the mass of the lightest sparticle
(LSP) from the CDM abundance in the universe is incompatible with
the data on ${\rm BR}(b\to s\gamma)$. Thus, the latter scheme is
not viable. \vskip0.10cm

The $\mu>0$ model possesses a number of other interesting features
too. The LSP is possibly detectable in the near future direct CDM
searches. The $\mu$ problem of MSSM is solved \cite{rsym} via a
Peccei-Quinn (PQ) symmetry \cite{pq} which also solves the strong
CP problem. Although the baryon ($B$) and lepton ($L$) numbers are
explicitly violated, the proton lifetime is considerably higher
than the present experimental limits. Furthermore, the same model
leads to a new shifted hybrid inflationary scenario \cite{jean2},
which avoids monopole overproduction at the end of inflation and
reproduces the results on the quadrupole anisotropy of the cosmic
microwave background radiation (CMBR) from the cosmic background
explorer (COBE) measurements \cite{cobe}. The relevant part of
inflation takes place at values of the inflaton field which are
not much smaller than the Planck scale and, thus, supergravity
(SUGRA) corrections could easily invalidate it. It is, however,
shown that inflation can be kept intact provided that an extra
gauge singlet is introduced and a specific form of the K\"{a}hler
potential is used. Although the SUGRA corrections are sizable, the
constraints from COBE can again be met by readjusting the values
of the parameters which were obtained with global SUSY.
\vskip0.10cm

The construction of the models is briefly reviewed in Sec.
\ref{model} and the resulting CMSSM is presented in Sec.
\ref{masspar}. The parameter space of the CMSSM is restricted in
Sec. \ref{results} taking into account a number of cosmological
and phenomenological requirements which are exhibited in Sec.
\ref{sec:pheno}. The deviation from YU is estimated in Sec.
\ref{delta}. Issues related to the direct LSP detection are
examinated  in Sec. \ref{phenoaa}. The resolution of the $\mu$
problem of MSSM and the stability of the proton are discussed in
Secs. \ref{sec:rsym} and \ref{sec:prot} respectively. The new
shifted inflationary scenario is outlined in Sec.~\ref{inflation}.
Finally, we summarize our conclusions in Sec. \ref{con}. Details
concerning the evaluation of the neutralino$-$nucleus cross
section are given in Appendix A.

\section{The Pati-Salam SUSY GUT Models}\label{model}

\subsection{The General Set-up}\label{gen}

We focus on a SUSY GUT model based on the PS gauge group $G_{\rm
PS}= SU(4)_{\rm c}\times SU(2)_L\times SU(2)_R$ described in
detail in Ref. \cite{jean} (see also Ref. \cite{talks}). The
representations and the transformation properties under $G_{\rm
PS}$ as well as the extra global charges of the various
superfields contained in the model are presented in Table 1, which
also contains the extra Higgs superfields required for
accommodating an adequate violation of YU (see below).

The left handed quark and lepton superfields of the $r$th
generation $(r=1,2,3)$ are accommodated in a pair of superfields
\beq \tilde F_r=\Fia~~\mbox{and}~~\tilde F^c_r=\Fca, \label{Fg}
\eeq
where tilde denotes transpose with respect to (w.r.t.) $SU(4)_{\rm
c}$. The gauge symmetry $G_{\rm PS}$ can be spontaneously broken
down to the standard model (SM) gauge group ($G_{\rm SM}$) through
the VEVs which the superfields
\beq \tilde H^c=\Hca~~\mbox{and}~~\tilde{\bar H}^c=\Hcb \label{Hg}
\eeq
acquire in their right handed neutrino directions $\nu^c_H$ and
$\bar\nu^c_H$. The model also contains a gauge singlet $S$ which
triggers the breaking of $G_{\rm PS}$, a $SU(4)_{\rm c}$ {\bf
6}-plet $G$ which gives \cite{leontaris} masses to the right
handed down quark type components of $H^c$, $\bar{H}^c$ and a pair
of gauge singlets $N$, $\bar{N}$ for solving \cite{rsym} the $\mu$
problem of the MSSM via a PQ symmetry (see Sec. \ref{sec:rsym}).
In addition to $G_{\rm PS}$, the model possesses two global $U(1)$
symmetries, namely a PQ and a R symmetry, as well as a discrete
$Z_2^{\rm mp}$ symmetry (`matter parity') under which $F$, $F^c$
change sign. Note that global continuous symmetries such as our PQ
and R symmetry can effectively arise \cite{laz1} from the rich
discrete symmetry groups encountered in many compactified string
theories (see e.g. Ref.~\cite{laz2}).

\renewcommand{\arraystretch}{1.2}

\begin{table}[!t]
\begin{center}
\begin{tabular}{|cccccc|}
\hline {\sc Super-}&{\sc Represe-}&{\sc
Trasfor-}&\multicolumn{3}{c|}{ {\sc Global} }
\\
\multicolumn{1}{|c}{\sc fields}&{\sc ntations} &{\sc mations}
&\multicolumn{3}{c|}{ {\sc Charges}}
\\
\multicolumn{1}{|c}{}&{\sc under $G_{\rm PS}$} &{\sc under $G_{\rm
PS}$}&{$R$} &{$PQ$} &{$Z^{\rm mp}_2$}
\\\hline \hline
\multicolumn{6}{|c|}{\sc Matter Superfields}
\\ \hline
{$F_r$} &{$({\bf 4, 2, 1})$}&$F_rU_L^{\dagger}U^\tr_{\rm c}$ &
$1/2$ & $-1$ &$1$
 \\
{$F^c_r$} & {$({\bf \bar 4, 1, 2})$}&$U_{\rm c}^\ast U_R^\ast
F^c_r$ &{ $1/2$ }&{$0$}&{$-1$}
\\ \hline
\multicolumn{6}{|c|}{\sc Higgs Superfields}
\\ \hline
{$H^c$} &{$({\bf \bar 4, 1, 2})$}& $U_{\rm c}^\ast U_R^\ast H^c$
&{$0$}&{$0$} & {$0$}
 \\
{$\bar H^c$}&$({\bf 4, 1, 2})$& $\bar{H}^cU^\tr_R U^\tr_{\rm
c}$&{$0$}&{$0$}&{$0$} \\
{$S$} & {$({\bf 1, 1, 1})$}&$S$ &$1$ &$0$ &$0$ \\
{$G$} & {$({\bf 6, 1, 1})$}&$U_{\rm c}GU^\tr_{\rm c}$ &$1$ &$0$
&$0$\\ \hline
{$h$} & {$({\bf 1, 2, 2})$}&$U_LhU^\tr_R$ &$0$ &$1$ &$0$
 \\ \hline
{$N$} &{$({\bf 1, 1, 1})$}& $N$ &{$1/2$}&{$-1$} & {$0$} \\
{$\bar N$}&$({\bf 1, 1, 1})$& $\bar N$&{$0$}&{$1$}&{$0$}
\\ \hline
\multicolumn{6}{|c|}{\sc Extra Higgs Superfields}
\\ \hline
$h^{\prime}$&{$({\bf 15, 2, 2})$} & $U_{\rm c}^\ast
U_Lh^{\prime}U^\tr_RU^\tr_{\rm c}$ & $0$ & $1$ &$0$
 \\
 $\bar h^{\prime}$&{$({\bf 15, 2, 2})$} &
$U_{\rm c}U_L\bar{h}^{\prime}U^\tr_RU_{\rm c}^\dagger$ & $1$ &
$-1$ &$0$
\\
$\phi$&$({\bf 15, 1, 3})$& $U_{\rm c}U_R\phi U_R^\dagger U_{\rm
c}^\dagger$ & $0$ & $0$ &$0$
\\
$\bar\phi$&{$({\bf 15, 1, 3})$} &$U_{\rm c}U_R\bar \phi
U_R^\dagger U_{\rm c}^\dagger$ & $1$ & $0$ &$0$
\\\hline
\end{tabular}
\end{center}
\caption{\sl\ftn The representations and transformations under
$G_{\rm PS}$ as well as the extra global charges of the
superfields of our model ($U_{\rm c}\in SU(4)_{\rm c},~U_{L}\in
SU(2)_{L},~U_{R}\in SU(2)_{R}$ and $\tr~,\dagger$ and $\ast$ stand
for the transpose, the hermitian conjugate and the complex
conjugate of a matrix respectively).}
\end{table}

In the simplest realization of this model \cite{leontaris}, the
electroweak doublets $H_1, H_2$ are exclusively contained in the
bidoublet superfield $h$, which can be written as
\beq h=\llgm{h_2&h_1}\rrgm=\bdh. \eeq
Under these circumstances, $H_i=h_i$ with $i=1,2$ and so the model
predicts YU at $M_{\rm GUT}$ (which is determined by the
requirement of the unification of the gauge coupling constants),
i.e.
\beq h_t:h_b:h_\tau=1:1:1, \label{exact} \eeq
since, after the breaking of $G_{\rm PS}$ to $G_{\rm SM}$, the
third family fermion masses originate from a unique term of the
underlying GUT as follows:
\beq \label{ffy} y_{33}\>F_3\langle h\rangle F_3^c= y_{33}\>\left(
- v_2\> t\; t^c + v_1\> b\> b^c + v_1\> \tau \>\tau^c
\right)+\cdots,~~\mbox{where}~~v_i=\langle H_i\rangle.\eeq

A moderate violation of YU can be accommodated by adding two new
Higgs superfields $h^{\prime}$ and $\bar h^\prime$ with
\beq h^\prime=\llgm{h^\prime_2&h^\prime_1}\rrgm~~\mbox{and}~~\bar
h^\prime=\llgm{\bar h^\prime_2&\bar h^\prime_1}\rrgm.
\label{hs}\eeq
In accordance with the global symmetries imposed (see Table 1),
$h^\prime$ can couple to $FF^c$ since $FF^c={\bf (15, 2, 2)}$,
whereas $\bar h^\prime$ can give mass to the color non-singlet
components of $h^\prime$ through a superpotential term $m\bar
h^\prime h^\prime~(m\sim M_{\rm GUT}\simeq2\times10^{16}~{\rm
GeV})$, which corresponds to the following $G_{\rm PS}$ invariant
quantity:
\beq m\Tr\left(\bar h^{\prime}\openep
h^{\prime\tr}\openep\right),~~\mbox{where}~~\openep=\llgm{0&1\cr-1
&0}\rrgm \label{mixa} \eeq
and \Tr\ denotes trace taken w.r.t. the $SU(4)_{\rm c}$ and
$SU(2)_L$ indices. Other important superpotential terms, which,
being non-renormalizable, are suppressed by the string scale
$M_{\rm S}\simeq5\times10^{17}~{\rm GeV}$, are
\bea \bar H^c H^c \bar h^\prime h=\Big({\bf 1},~{\bf 1},~{\bf
1}\times{\bf 1}+{\bf 3}\times{\bf 3}\Big), \label{nonren} \eea
since
\begin{eqnarray} && \nonumber \bar H^c H^c =(
{\bf 4,1,2})({\bf\bar 4,1,2})=({\bf 15, 1,
1+3})+\cdots,\\ \nonumber && \bar h^\prime h=({\bf 15,2,2})({\bf
1,2,2})=({\bf 15, 1, 1+3})+\cdots. \end{eqnarray}
We see that, in Eq. (\ref{nonren}), there are two independent
couplings: a coupling between the $SU(2)_R$ singlets in $\bar H^c
H^c$ and $\bar h^\prime h$, $(\bar H^c H^c)_{\bf 1}\bar h^\prime
h$, and a coupling between their triplets, $(\bar H^c
H^c)_{\bf3}\bar h^\prime h$. As it turns out, the singlet coupling
provides us with an adequate deviation from YU for $\mu<0$. On the
other hand, $\mu>0$ requires a stronger deviation.

\subsection{The {\boldmath $\mu>0$} Case}\label{mupos}

The necessary deviation from YU for $\mu>0$ can be obtained by a
further enlargement of the Higgs sector so that contributions to
the coupling between $h$ and $\bar h^\prime$ from renormalizable
terms are also allowed. To this end, we introduce two additional
Higgs superfields $\phi,~\bar \phi$. The superfield $\bar\phi$ is
aimed to give superheavy masses to the color non-singlets in
$\phi$ through a term $\bar\phi\phi$, whose coefficient is of
order $M_{\rm GUT}$. The superfield $\phi$, on the other hand,
yields the unsuppressed coupling $\lambda_{\bf 3} \phi \bar
h^\prime h$ (with $\lambda_{\bf3}$ being a dimensionless
constant), which overshadows the coupling $(\bar H^c
H^c)_{\bf3}\bar h^\prime h$ and corresponds to the $G_{\rm PS}$
invariant term
\beq \lambda_{\bf 3}\Tr\left(\bar h^{\prime} \openep\phi h^\tr
\openep \right).\label{expmixp} \eeq

During the spontaneous breaking of $G_{\rm PS}$ to $G_{\rm SM}$,
$\phi$ acquires VEV in the SM singlet direction. Therefore,
\beq \langle\phi\rangle=v_\phi
\left(T^{15},1,\frac{\sigma_3}{\sqrt{2}} \right),\label{vevs}\eeq
where $v_{\phi}\sim M_{\rm GUT}$ and the structure of
$\langle\phi\rangle$ w.r.t. $G_{\rm PS}$ is shown with
\beq T^{15}= \frac{1}{2\sqrt{3}}\>{\sf
diag}\Big(1,1,1,-3\Big)~~\mbox{and}~~\sigma_{3}={\sf
diag}\Big(1,-1\Big).\eeq
Expanding the superfields in Eq. (\ref{hs}) as linear combination
of the fifteen generators $T^a$ of $SU(4)_{\rm c}$ with the
normalization $\Tr(T^aT^b)=\delta^{ab}$ and denoting the SM
singlet components with the superfield symbol, we can easily
establish the following identities:
\bea && \Tr\left(\bar h^{\prime} \openep h^{\prime\tr}
\openep\right)=\bar h^{\prime\tr}_1\openep h^\prime_2 +
h^{\prime\tr}_1\openep \bar h^\prime_2+\cdots,\label{idnta}
\\ && \Tr\left(\bar h^{\prime}
\openep\left(T^{15},1,\sigma_{3}\right)h^\tr
\openep\right)=\left(\bar h^{\prime\tr}_1\openep h_2 -
h^\tr_1\openep\bar h^\prime_2\right), \label{idnt}\eea
where the notation of Eq. (\ref{vevs}) has been applied.

Inserting Eq. (\ref{vevs}) in Eq. (\ref{expmixp}), employing Eqs.
(\ref{idnta}) and (\ref{idnt}), and collecting Eqs.~(\ref{mixa})
and (\ref{expmixp}) together, we get the mass terms
\beq m\bar h^{\prime\tr}_1\openep\left(h^{\prime}_2+
\alpha_2h_2\right)+m\left(h^{\prime \tr}_1+\alpha_1
h^\tr_1\right)\openep\bar{h}^{\prime}_2, \label{superheavy}\eeq
where the mixing effects are included in the following
coefficients:
\beq \alpha_{1}=-\alpha_{2}= -\lambda_{\bf 3}v_\phi/\sqrt{2}m.
\label{alphas} \eeq

\subsection{The {\boldmath $\mu<0$} Case}\label{muneg}

In the $\mu<0$ case, an adequate violation of YU can be achieved
(without the inclusion of $\phi$ and $\bar\phi$) predominantly via
the non-renormalizable $SU(2)_R$ singlet coupling $\lambda_{\bf 1}
(\bar H^c \bar H^c)_{\bf 1} h^\prime h/M_{\rm S}$ (with
$\lambda_{\bf1}$ being a dimensionless constant), which
corresponds to the $G_{\rm PS}$ invariant term
\beq \frac{\lambda_{\bf 1}}{M_{\rm S}}\Tr\left(\bar
h^{\prime}\openep(\bar H^{c\tr} H^{c\tr})_{\bf1} h^\tr
\openep\right).\label{sigl}\eeq
During the spontaneous breaking of $G_{\rm PS}$ to $G_{\rm SM}$,
$H^c$ and $\bar H^c$ acquire VEVs in the SM singlet direction.
Therefore,
\beq \left(\langle\bar H^{c\tr}\rangle\langle
H^{c\tr}\rangle\right)_{{\bf1}}=v^2_{H^c}\left(T_{H^c},1,\frac{\sigma_{0}}{2}
\right),\label{vevH}\eeq
where $v_{H^c}\sim M_{\rm GUT}$ and the notation of Eq.
(\ref{vevs}) has been applied with
\beq T_{H^c}={\sf
diag}\Big(0,0,0,1\Big)~\quad\mbox{and}\quad~\sigma_{0}={\sf
diag}\Big(1,1\Big).\eeq
Expanding the superfields in Eq. (\ref{hs}) as we did in deriving
Eq. (\ref{idnt}), we obtain the following identity:
\beq \Tr\left(\bar h^{\prime}
\openep\left(T_{H^c},1,\sigma_{0}\right)h^\tr
\openep\right)=-\sqrt{3}\left(\bar h^{\prime\tr}_1\openep h_2 +
h^\tr_1\openep\bar h^\prime_2\right)/2. \label{idntH}\eeq
Inserting Eq. (\ref{vevH}) in Eq. (\ref{sigl}), employing
Eqs.~(\ref{idnta}) and (\ref{idntH}), and collecting
Eqs.~(\ref{mixa}) and (\ref{sigl}) together, we end up again with
Eq. (\ref{superheavy}) but with
\beq \alpha_{1}=\alpha_{2}= -\frac{\sqrt{3}\lambda_{\bf
1}}{4M_{\rm S}m}v^2_{H^c}, \label{alphasn} \eeq
which are suppressed by $M_{\rm GUT}/M_{\rm S}$.

\subsection{Yukawa Quasi-Unification Conditions}\label{Yqucs}

It is obvious from Eq. (\ref{superheavy}) that we obtain two
combinations of superheavy massive fields
\bea\bar{h}^{\prime}_1,~H^{\prime}_1~~\mbox{and}~~
\bar{h}^{\prime}_2,~H^{\prime}_2,~~\mbox{where}~~
H^{\prime}_{i}=\frac{h^{\prime}_{i}+\alpha_{i}h_{i}}
{\sqrt{1+|\alpha_{i}|^2}},~i=1,2. \label{shs}\eea
The electroweak doublets $H_i$, which remain massless at the GUT
scale, are orthogonal to the $H^{\prime}_{i}$ directions:
\beq H_i=\frac{-\alpha_i^*h^{\prime}_i+h_i}
{\sqrt{1+|\alpha_i|^2}} \cdot\label{elws}\eeq
\par
Solving Eqs. (\ref{shs}) and (\ref{elws}) w.r.t. $h_i$ and
$h^\prime_i$, we obtain
\beq h_i=\frac{H_i+\alpha^*_iH^{\prime}_i}{\sqrt{1+|\alpha_i|^2}}
~~\mbox{and}~~h^\prime_i=\frac{-\alpha_iH_i+H^{\prime}_i}
{\sqrt{1+|\alpha_i|^2}}\cdot~~~\eeq
Consequently, the third family fermion masses are now generated by
the following terms (compare with Eq. (\ref{ffy})):
$$ y_{33}\>F_3\langle h\rangle F_3^c+2y^\prime_{33}\>F_3\langle
h^\prime\rangle F_3^c=$$
$$y_{33}\>\left(-\frac{1-\rho\alpha_2/\sqrt{3}}{\sqrt{1+|\alpha_2|^2}}\>v_2\>
t\;t^c+\frac{1-\rho\alpha_1/\sqrt{3}}{\sqrt{1+|\alpha_1|^2}}\>v_1\>b\>b^c
+ \frac{1+\sqrt{3}\rho\alpha_1}{\sqrt{1+|\alpha_1|^2}}\>v_1\>\tau
\>\tau^c\right)+\cdots$$
where $\rho=y^\prime_{33}/y_{33}$ with $0<\rho<1$ and the color
non-singlet components of $h^\prime_i$ are included in the
ellipsis. The fields $H^\prime_i$, being superheavy, contribute
neither to the running of renormalization group equations (RGEs)
nor to the masses of the fermions. Consequently, the asymptotic
exact YU in Eq. (\ref{exact}) can be replaced by a set of
asymptotic Yukawa quasi-unification conditions (YQUCs):
\bea h_t:h_b:h_\tau= \left|\frac{1-\rho\alpha_2/\sqrt{3}}
{\sqrt{1+|\alpha_2|^2}}\right|:
\left|\frac{1-\rho\alpha_1/\sqrt{3}}
{\sqrt{1+|\alpha_1|^2}}\right|:
\left|\frac{1+\sqrt{3}\rho\alpha_1}
{\sqrt{1+|\alpha_1|^2}}\right|\cdot\label{qg}\eea
Substituting Eqs. (\ref{alphas}) and (\ref{alphasn}) into Eq.
(\ref{qg}), we obtain
\beq \kern-4.pt h_t:h_b:h_\tau=\left\{\matrix{
%\begin{array}{llll}
(1+c):(1-c):(1+3c)\hfill &
\mbox{with}&0<c<1 &\mbox{for}~\mu>0 \cr
(1-c):(1-c):(1+3c)\hfill &
\mbox{with}&-1/3<c<0&\mbox{for}~\mu<0\cr}
%\end{array}
\right..\label{minimal} \eeq
For simplicity, we restricted ourselves to real values of $c$
only, where $c=\rho\alpha_1/\sqrt{3}$.

The deviation from YU can be estimated by defining the following
relative splittings:
\beq \delta
h_{b[\tau]}=\frac{h_{b[\tau]}-h_t}{h_t}=\left\{\matrix{
%\begin{array}{rl}
-2c/(1+c)~[2c/(1+c)]\hfill  & \mbox{for}~\mu>0\hfill \cr
0~[4c/(1+c)]\hfill & \mbox{for}~\mu<0 \hfill \cr}
%\end{array}
\right.. \label{dhdef} \eeq

\section{The resulting CMSSM} \label{masspar}

%%%%%%%%%%%%%%%%%%%%%%%%%%%%%%%%%%%%%%%%%%%%%%%%%%%%%%%%%%%%%%%%%%%%
\begin{figure}[!ht]
\centerline{\epsfig{file=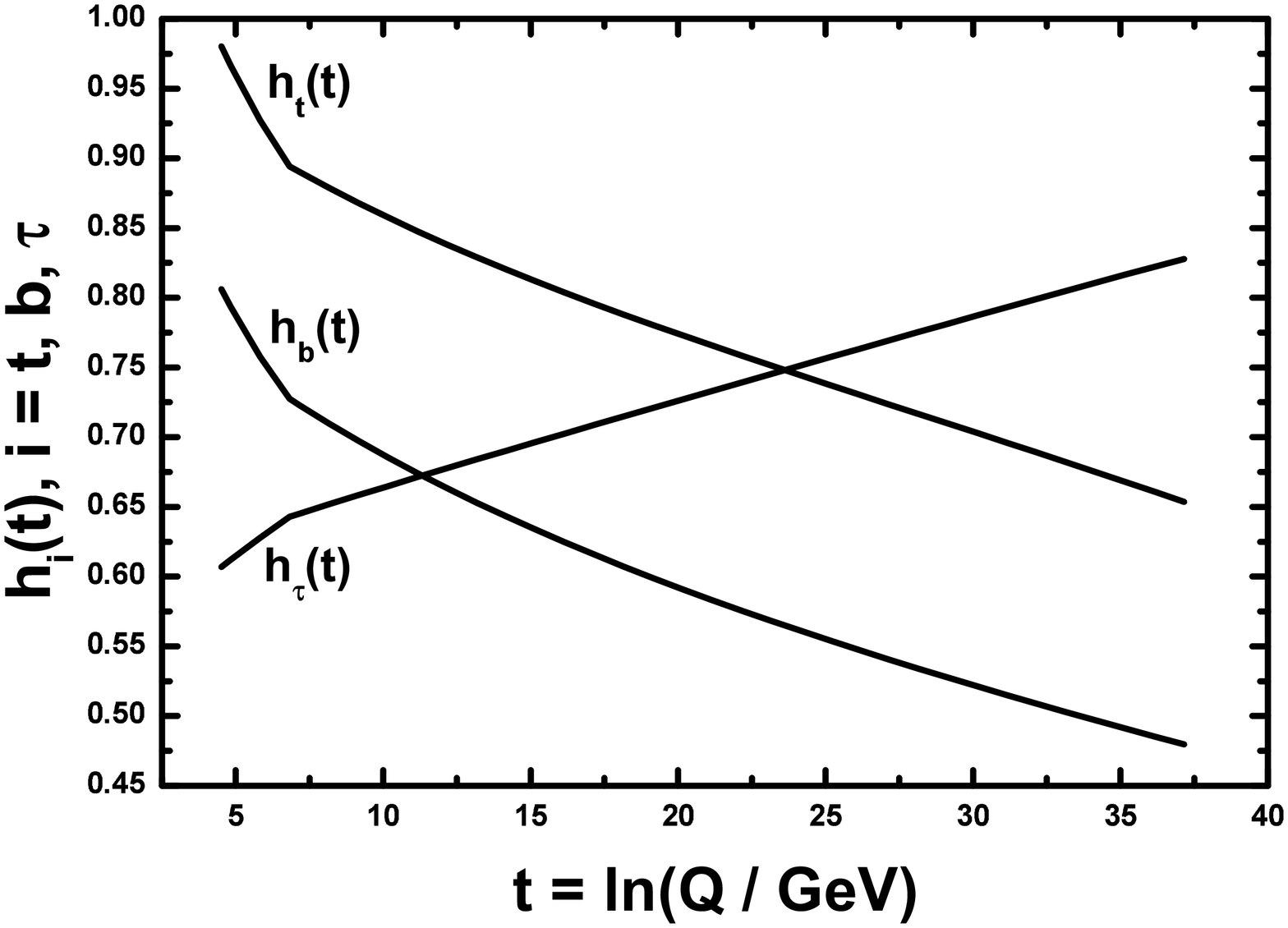,angle=-90,width=11cm}} \hfill
\caption{\sl\ftn A running of the Yukawa coupling constants from
$Q=M_{\rm GUT}$ to $Q=M_Z$ for $m_b(M_Z)=2.888~{\rm GeV}$,
$\mu>0$, $m_{\rm LSP}=200~{\rm GeV}$ and $\Delta_{\tilde\tau_2}=1$
corresponding to $\tan\beta\simeq 58$.}\label{yukp}
\end{figure}
%%%%%%%%%%%%%%%%%%%%%%%%%%%%%%%%%%%%%%%%%

%%%%%%%%%%%%%%%%%%%%%%%%%%%%%%%%%%%%%%%%%%%%%%%%%%%%%%%%%%%%%%%%%%%%
\begin{figure}[!ht]
%\vspace*{-0.4cm}
\centerline{\epsfig{file=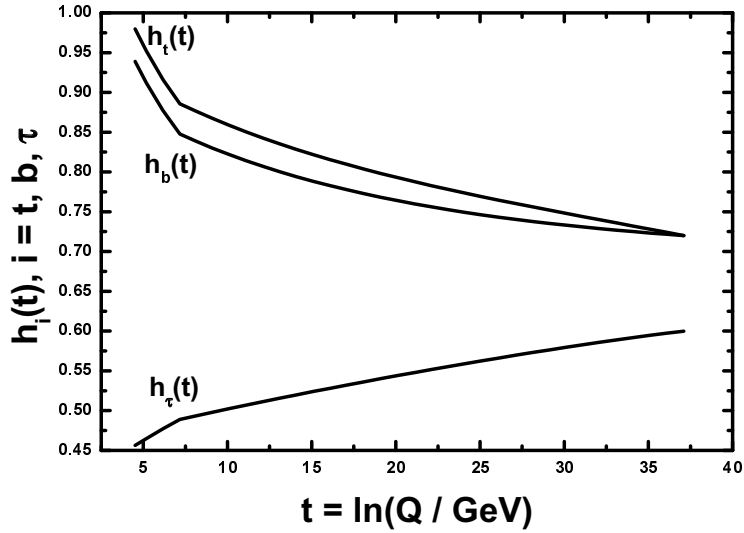,angle=-90,width=11cm}} \hfill
\caption{\sl\ftn A running of the Yukawa coupling constants from
$Q=M_{\rm GUT}$ to $Q=M_Z$ for $m_b(M_Z)=2.888~{\rm GeV}$,
$\mu<0$, $m_{\rm LSP}=350~{\rm GeV}$ and $\Delta_{\tilde\tau_2}=0$
corresponding to $\tan\beta\simeq 47$.}\label{yukn}
\end{figure}
%%%%%%%%%%%%%%%%%%%%%%%%%%%%%%%%%%%%%%%%%

Below $M_{\rm GUT}$, the particle content of our models reduces to
this of MSSM (modulo SM singlets). We assume universal soft SUSY
breaking scalar masses $m_0$, gaugino masses $M_{1/2}$ and
trilinear scalar couplings $A_0$ at $M_{\rm GUT}$. Therefore, the
resulting MSSM is the so-called CMSSM \cite{Cmssm} supplemented by
a suitable YQUC from the set in Eq. (\ref{minimal}). Let us
emphasize that the specific relations between $\alpha_1$ and
$\alpha_2$ given in Eq. (\ref{alphas})~[Eq.~(\ref{alphasn})] for
$\mu>0~[\mu<0]$ ensure a SUSY spectrum which leads to successful
radiative electroweak symmetry breaking (REWSB) and a neutral LSP
in a large fraction of the parametric space in the framework of
the CMSSM (and not in a general version of MSSM).

We integrate the two-loop RGEs for the gauge and Yukawa coupling
constants and the one-loop ones for the soft SUSY breaking terms
between $M_{\rm GUT}$ and a common SUSY threshold $M_{\rm SUSY}
\simeq(m_{\tilde t_1}m_{\tilde t_2})^{1/2}$ ($\tilde t_{1,2}$ are
the stop mass eigenstates) determined in consistency with the SUSY
spectrum. At $M_{\rm SUSY}$, we impose REWSB, evaluate the SUSY
spectrum and incorporate the SUSY corrections to the $b$ and
$\tau$ mass \cite{pierce, susy, king}. The corrections to $m_\tau$
(almost 4$\%$) lead \cite{cdm} to a small decrease [increase] of
$\tan\beta$ for $\mu>0~[\mu<0]$. From $M_{\rm SUSY}$ to $M_Z$, the
running of gauge and Yukawa coupling constants is continued using
the SM RGEs.

For presentation purposes, $M_{1/2}$ and $m_0$ can be replaced
\cite{cdm} by the LSP mass, $m_{\rm LSP}$, and the relative mass
splitting, $\Delta_{\tilde\tau_2}$, between the LSP and the
lightest stau $\tilde\tau_2$. For simplicity, we restrict this
presentation to the $A_0=0$ case (for $A_0\neq0$, see
Refs.~\cite{qcdm, mario}). So, our free input parameters are
$${\rm sign}(\mu),~m_{\rm LSP},~\Delta_{\tilde\tau_2},
~~\mbox{with}~~\Delta_{\tilde\tau_2}=(m_{\tilde\tau_2}-m_{\rm
LSP})/m_{\rm LSP}.$$

For any given $m_b(M_Z)$ in the range in Eq. (\ref{mbrg}) with
fixed masses for the top quark $m_t(m_t)=166~{\rm GeV}$ and the
tau lepton $m_\tau(M_Z)=1.746~{\rm GeV}$, we can determine the
parameters $c$ and $\tan\beta$ at $M_{\rm SUSY}$ so that the
corresponding YQUC in Eq. (\ref{minimal}) is satisfied.
Consequently, for any given $m_b(M_Z)$, a prediction for
$\tan\beta$ can be made, in contrast to the original version of
CMSSM  \cite{Cmssm}.

In Fig. \ref{yukp} [Fig. \ref{yukn}], we present a Yukawa coupling
constant running from $M_{\rm GUT}$ to $M_Z$ for the central value
of $m_b(M_Z)=2.888~{\rm GeV}$. At $M_{\rm GUT}$, we apply Eq.
(\ref{minimal}) with $c=0.154~[c=-0.043]$ (corresponding to
$\tan\beta\simeq58~[\tan\beta\simeq47]$ at $M_{\rm SUSY}$) for
$\mu>0~[\mu<0]$, $m_{\rm LSP}=200~{\rm GeV}~[m_{\rm LSP}=350~{\rm
GeV}]$ and $\Delta_{\tilde\tau_2}=1~[\Delta_{\tilde\tau_2}=0]$.
The kinks on the various curves correspond to the point where the
MSSM RGEs are replaced by the SM ones. We observe that, for
$\mu>0$, $h_\tau$ and $h_b$ split from $h_t$ by the same amount
but in opposite directions with $h_b$ becoming smaller than $h_t$
(see Fig. \ref{yukp}), while, for $\mu<0$, $h_t$ and $h_b$ remain
unified (see Fig. \ref{yukn}).

%%%%%%%%%%%%%%%%%%%%%%%%%%%%%%%%%%%%%%%%%%%%%%%%%%%%%%%%%%%%%%%%%%%%
\begin{figure}[!th]
\centerline{\epsfig{file=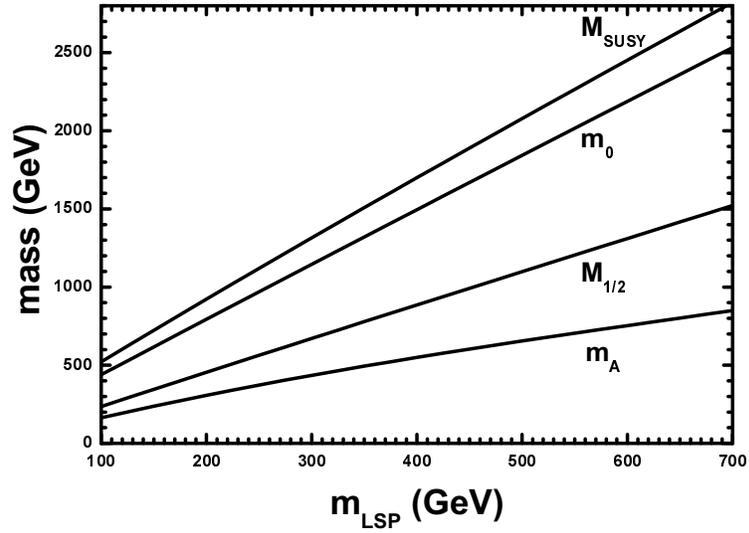,angle=-90,width=11cm}} \hfill
\caption{\sl\ftn The mass parameters $m_A$, $m_0$, $M_{1/2}$ and
$M_{\rm SUSY}$ versus $m_{\rm LSP}$ for $\mu>0$, $m_b(M_Z)=2.888$
$~{\rm GeV}$ and $\Delta_{\tilde\tau_2}=1$.}\label{Mpx}
\end{figure}
%%%%%%%%%%%%%%%%%%%%%%%%%%%%%%%%%%%%%%%%%(see Sec. \ref{rpos} [\ref{rmuneg}])
%%%%%%%%%%%%%%%%%%%%%%%%%%%%%%%%%%%%%%%%%%%%%%%%%%%%%%%%%%%%%%%%%%%%
\begin{figure}[!ht]
\centerline{\epsfig{file=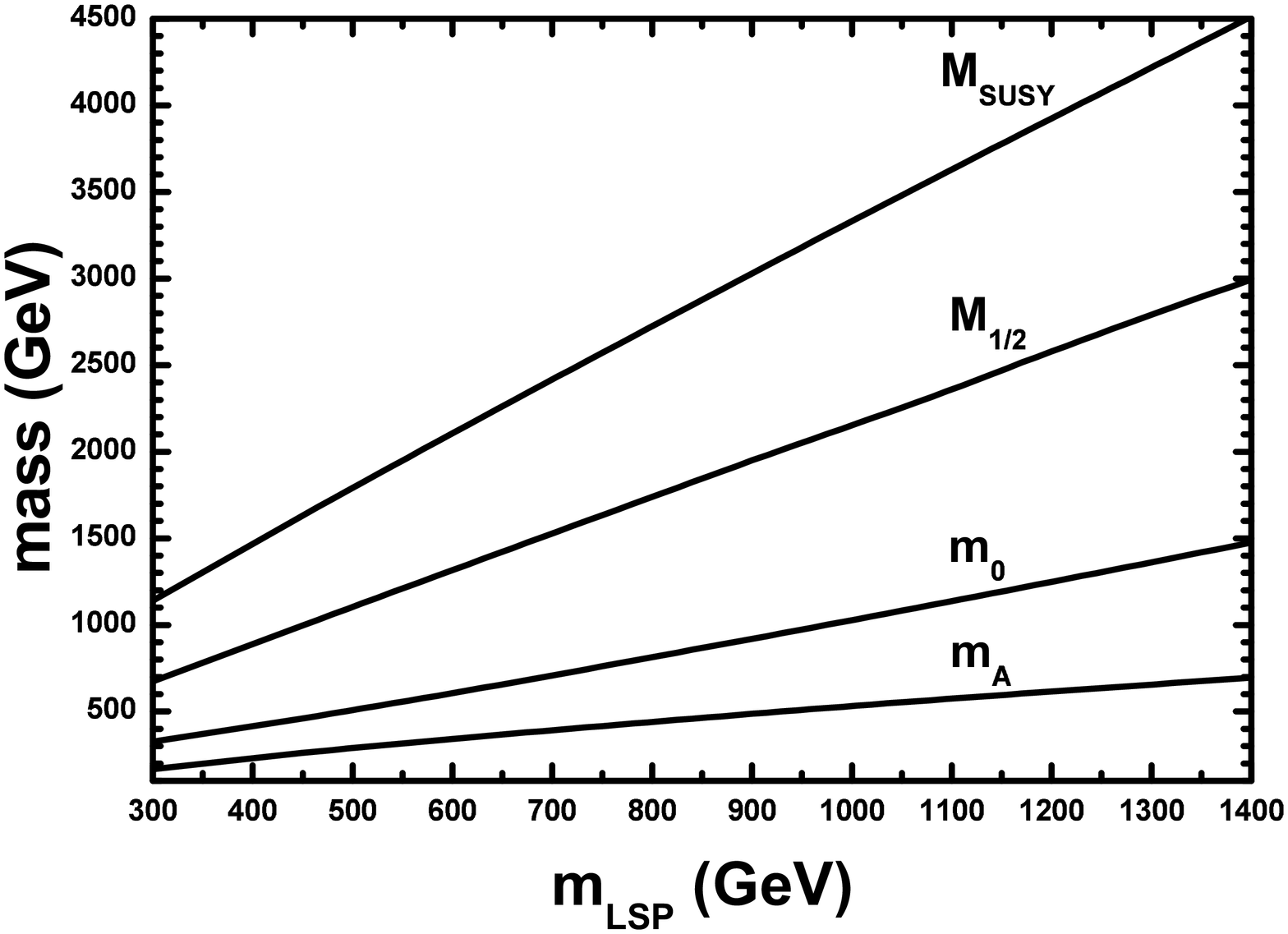,angle=-90,width=11cm}} \hfill
\caption{\sl\ftn The mass parameters $m_A$, $m_0$, $M_{1/2}$ and
$M_{\rm SUSY}$ versus $m_{\rm LSP}$ for $\mu<0$, $m_b(M_Z)=2.888$
$~{\rm GeV}$ and $\Delta_{\tilde\tau_2}=0$.}\label{Mnx}
\end{figure}
%%%%%%%%%%%%%%%%%%%%%%%%%%%%%%%%%%%%%%%%%

In Fig. \ref{Mpx}~[Fig. \ref{Mnx}], we present the values of the
mass parameters $m_A$, $m_0$, $M_{1/2}$ and $M_{\rm SUSY}$ versus
$m_{\rm LSP}$ for $\mu>0~[\mu<0]$,
$\Delta_{\tilde\tau_2}=1~[\Delta_{\tilde\tau_2}=0]$ (see
Sec.~\ref{rpos} [Sec.~\ref{rmuneg}]) and with $m_b(M_Z)=2.888$
GeV. We observe that $m_0\gg M_{1/2}~ [m_0\ll M_{1/2}]$ and
$m_A\lesssim 2m_{\rm LSP}~[m_A\ll 2m_{\rm LSP}]$ for
$\mu>0~[\mu<0]$. The relation between $m_{\rm LSP}$ and $m_A$ or
$m_{\tilde\tau_2}$ is crucial for the reduction of the LSP relic
abundance (see Sec. \ref{phenoa}).

\section{Cosmological and Phenomenological Constraints} \label{sec:pheno}

The parameter space of our models can be restricted by using a
number of phenomenological and cosmological constraints. We will
now briefly discuss these requirements (for similar recent
analyses, see Refs. \cite{baery, nath, su5b, spanos}).

\subsection{Cold Dark Matter Considerations}
\label{phenoa}

According to the WMAP results \cite{wmap}, the $95\%$ c.l. range
for the CDM abundance is
\beq \Omega_{\rm CDM}h^2=0.1126_{-0.0181}^{+0.0161}.
\label{cdmba}\eeq
In the context of the CMSSM, the LSP can be the lightest
neutralino $\tilde\chi$. It naturally arises \cite{goldberg} as a
CDM candidate. We require its relic abundance, $\Omega_{\rm
LSP}h^2$, not to exceed the upper bound derived from Eq.
(\ref{cdmba}) (the lower bound is not considered since other
production mechanisms of LSPs may be present too \cite{scn, scna}
and/or other CDM candidates \cite{axion, axino} may also
contribute to $\Omega_{\rm CDM}h^2$):
\beq \Omega_{\rm LSP}h^2\lesssim0.13.\label{cdmb}\eeq
For each sign of $\mu$, an upper bound on $m_{\rm LSP}$ (or
$m_{\tilde\chi}$) can be derived from Eq.~(\ref{cdmb}). We
calculate $\Omega_{\rm LSP}h^2$ using the publicly available code
{\tt micrOMEGAs} \cite{micro} (not the latest version
\cite{micron}). This includes accurately thermally averaged exact
tree-level cross sections of all possible (co)annihilation
processes, treats poles properly and uses one-loop QCD (not SUSY
QCD \cite{micron, pallis}) corrections to the Higgs decay widths
and couplings to fermions \cite{width}. Good agreement between
this code and other independent calculations of $\Omega_{\rm
LSP}h^2$ which include the $A$-pole effect and neutralino-stau
coannihilation is recently reported in the first paper of Ref.
\cite{qcdm} and in Ref. \cite{pallis}.

In most of the parameter space of the CMSSM, the LSP is an almost
pure bino and $\Omega_{\rm LSP}h^2$ increases with $m_{\rm LSP}$.
Therefore, Eq. (\ref{cdmb}) sets a very stringent upper limit on
the LSP mass \cite{munoz}. However, as pointed out in Refs.
\cite{lah, ellis}, a substantial reduction of $\Omega_{\rm
LSP}h^2$ can be achieved in some regions of the parameter space
thanks to two mechanisms: the $A$-pole effect and the bino-slepton
coannihilations. The first is activated for $\tan\beta>40~
[\tan\beta\simeq 30-35]$ for $\mu>0~[\mu<0]$, where the presence
of a resonance ($2m_{\rm LSP}\simeq m_A$) in the LSP annihilation
via a $s$-channel exchange of an $A$-boson is possible. On the
other hand, coannihilations can be activated for every $\tan\beta$
and both signs of $\mu$, but this requires a proximity between the
masses of the LSP and the next-to-LSP, which turns out to be the
$\tilde\tau_2$ for $\tan\beta>10$ \cite{ellis} and not too large
values of $A_0$ \cite{drees} or $m_0$ \cite{darkn}. For fixed
$m_{\rm LSP}$, $\Omega_{\rm LSP}h^2$ decreases with
$\Delta_{\tilde\tau_2}$, since the $\tilde\chi-\tilde\tau_2$
coannihilations become more efficient. So the CDM criterion can be
used for restricting $\Delta_{\tilde\tau_2}$.

Our models give us the opportunity to discuss how both these
reduction mechanisms operate. As we noticed from Figs. \ref{Mpx}
and \ref{Mnx}, for $\mu>0$, there is a significant region
dominated by the $A$-pole effect, while, for $\mu<0$, the
$\tilde\chi-\tilde\tau_2$ coannihilation is the only available
reduction mechanism.

\subsection{Branching Ratio of {\boldmath $b\to s\gamma$}}\label{phenog}

Taking into account the recent experimental results \cite{cleo} on
${\rm BR}(b\rightarrow s\gamma)$ and combining \cite{qcdm}
appropriately the various experimental and theoretical errors
involved, we obtain the following $95\%$ c.l. range:
\beq \mbox{\sf a)}~1.9\times 10^{-4}\lesssim {\rm BR}(b\rightarrow
s\gamma)~~~\mbox{and}~~~
\mbox{\sf b)}~{\rm BR}(b\rightarrow s\gamma)\lesssim 4.6 \times
10^{-4}. \label{bsgb} \eeq
We compute ${\rm BR}(b\rightarrow s\gamma)$ by using an updated
version of the relevant calculation contained in the {\tt
micrOMEGAs} package \cite{micro}.  In this code, the SM
contribution is calculated using the formalism of Ref.
\cite{kagan}. The charged Higgs boson, $H^\pm$, contribution is
evaluated by including the next-to-leading order (NLO) QCD
corrections from Ref. \cite{nlohiggs} and the $\tan\beta$ enhanced
contributions from Ref. \cite{nlosusy}. The dominant SUSY
contribution, ${\rm BR}(b\rightarrow s\gamma)|_{\rm SUSY}$,
includes resummed NLO SUSY QCD corrections from Ref.
\cite{nlosusy}, which hold for large $\tan\beta$.  The $H^\pm$
contribution interferes constructively with the SM contribution,
whereas ${\rm BR}(b\rightarrow s\gamma)|_{\rm SUSY}$ interferes
destructively [constructively] with the other two contributions
for $\mu>0~[\mu<0]$. Although the improvements of Ref.
\cite{isidori} are not included in this routine, their impact is
not important for $\mu>0$, whereas, for $\mu<0$, they are not
expected to change essentially our conclusions. The SM plus
$H^\pm$ contribution and the ${\rm BR}(b\rightarrow s\gamma)|_{\rm
SUSY}$ decrease as $m_{\rm LSP}$ increases and so a lower bound on
$m_{\rm LSP}$ can be derived from Eq.~(\ref{bsgb}{\sf a})
[Eq.~(\ref{bsgb}{\sf b})] for $\mu>0~[\mu<0]$ with the bound for
$\mu<0$ being much more restrictive.

\subsection{Muon Anomalous Magnetic Moment}\label{phenoc}

The deviation, $\delta a_\mu$, of the measured value of the muon
anomalous magnetic moment, $a_\mu$, from its predicted value in
the SM, $a^{\rm SM}_\mu$, can be attributed to SUSY contributions
arising from chargino-sneutrino and neutralino-smuon loops. The
quantity $\delta a_\mu$ is calculated by using the {\tt
micrOMEGAs} routine based on the formulas of Ref. \cite{gmuon}.
The absolute value of the result decreases as $m_{\rm LSP}$
increases and its sign is positive [negative] for $\mu>0~[\mu<0]$.

On the other hand, the calculation of $a^{\rm SM}_\mu$ is not yet
stabilized mainly because of the instability of the hadronic
vacuum polarization contribution. According to the  most
up-to-date evaluation of this contribution in Ref. \cite{davier},
there is still a discrepancy between the findings based on the
$e^+e^-$ annihilation data and the ones based on the $\tau$-decay
data. Taking into account these results and the recently announced
experimental measurements \cite{muon} on $a_\mu$, we impose the
following $95\%$ c.l. ranges:
\bea\mbox{\sf a)}~-0.53\times10^{-10}\lesssim\delta
a_\mu&~\mbox{and}~&\mbox{\sf b)}~\delta a_\mu\lesssim 44.7\times
10^{-10},~~\mbox{$e^+e^-$-based};\label{g2e}
\\\vspace*{19pt}
\mbox{\sf a)}~-13.5\times10^{-10}\lesssim\delta
a_\mu&~\mbox{and}~&\mbox{\sf b)}~\delta a_\mu\lesssim28.4\times
10^{-10},~~\mbox{$\tau$-based}.\label{g2t} \eea
A lower bound on $m_{\rm LSP}$ can be derived from Eq.
(\ref{g2e}{\sf b})~[Eq.~(\ref{g2t}{\sf a})] for $\mu>0~[\mu<0]$.
Although the $\mu<0$ case can not be excluded \cite{gmuon2}, it is
considered as quite disfavored \cite{narison} because of the poor
$\tau$-decay data.

\subsection{Collider Bounds}
\label{phenod}

For our analysis, the only relevant collider bound is the $95\%$
c.l. LEP bound \cite{higgs} on the lightest CP-even neutral Higgs
boson, $h$, mass
\beq m_h\gtrsim114.4~{\rm GeV},\label{mhb} \eeq
which gives a [almost always the absolute] lower bound on $m_{\rm
LSP}$ for $\mu<0$ $[\mu>0]$. The SUSY contributions to $m_h$ are
calculated at two loops by employing the program {\tt
FeynHiggsFast} \cite{fh} contained in the {\tt micrOMEGAs} package
\cite{micro}.

\section{Restrictions on the SUSY Parameters} \label{results}

Applying the cosmological and phenomenological requirements given
in Sec.~\ref{sec:pheno}, we delineate on the $m_{\rm
LSP}-\Delta_{\tilde\tau_2}$ plane the allowed parameter space of
our models  in Secs. \ref{rpos} and \ref{rmuneg} for $\mu>0$ and
$\mu<0$ respectively. For simplicity, $\alpha_s(M_Z)$ is fixed to
its central experimental value (equal to 0.1185) throughout our
calculation. Note that allowing it to vary in its $95\%$ c.l.
experimental range, $0.1145-0.1225$, the range of $m_b(M_Z)$ in
Eq. (\ref{mbrg}) would be slightly widened. This would lead to a
sizeable enlargement of the allowed area for $\mu>0$ due to the
sensitivity of $\Omega_{\rm LSP}h^2$ to the $b$-quark mass (see
Sec. \ref{rpos}). On the contrary, our results for $\mu<0$ would
be essentially unaffected, since the LSP annihilation to $b\bar b$
via an $A$-boson exchange gives a subdominant contribution to
$\Omega_{\rm LSP}h^2$ (see Sec. \ref{rmuneg}).

\subsection{The {\boldmath $\mu>0$} Case}\label{rpos}

%%%%%%%%%%%%%%%%%%%%%%%%%%%%%%%%%%%%%%%%%%%%%%%%%%%%%%%%%%%%%%%%%%%%
\begin{figure}[t]
\centerline{\epsfig{file=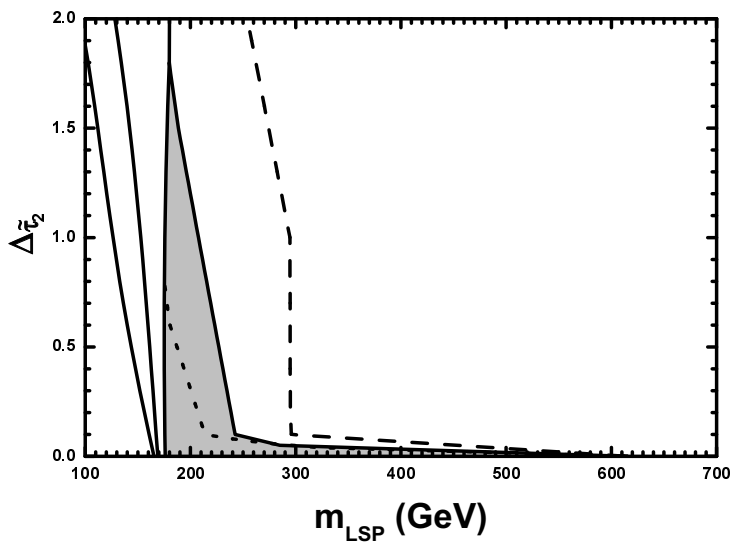,angle=-90,width=11.cm}} \hfill
\caption{\sl\ftn Restrictions on the $m_{\rm
LSP}-\Delta_{\tilde\tau_2}$ plane for $\mu>0$. From left to right,
the solid lines depict the lower bounds on $m_{\rm LSP}$ from Eq.
(\ref{g2e}{\sf b}), (\ref{bsgb}{\sf a}), (\ref{mhb}) and the upper
bound on $m_{\rm LSP}$ from Eq.~(\ref{cdmb}) for
$m_b(M_Z)=2.888~{\rm GeV}$. The dashed [dotted] line depicts the
bound on $m_{\rm LSP}$ from Eq.~(\ref{cdmb}) for
$m_b(M_Z)=2.684~{\rm GeV}~[3.092~{\rm GeV}]$. The allowed area for
$m_b(M_Z)=2.888~{\rm GeV}$ is shaded.} \label{figa}
\end{figure}
%%%%%%%%%%%%%%%%%%%%%%%%%%%%%%%%%%%%%%%%%

The restrictions on the $m_{\rm LSP}-\Delta_{\tilde\tau_2}$ plane
with $m_b(M_Z)=2.888~{\rm GeV}$ are shown in Fig. \ref{figa} as
solid lines, while the upper bound on $m_{\rm LSP}$ from Eq.
(\ref{cdmb})  for $m_b(M_Z)=2.684~{\rm GeV}~[m_b(M_Z)=3.092~{\rm
GeV}$] is depicted by a dashed [dotted] line. Needless to say that
the constraints from Eqs. (\ref{bsgb}{\sf b}) and (\ref{g2t}{\sf
a}) do not restrict the parameters, since they are always
satisfied for $\mu>0$.

\begin{figure}[!th]
\centerline{\epsfig{file=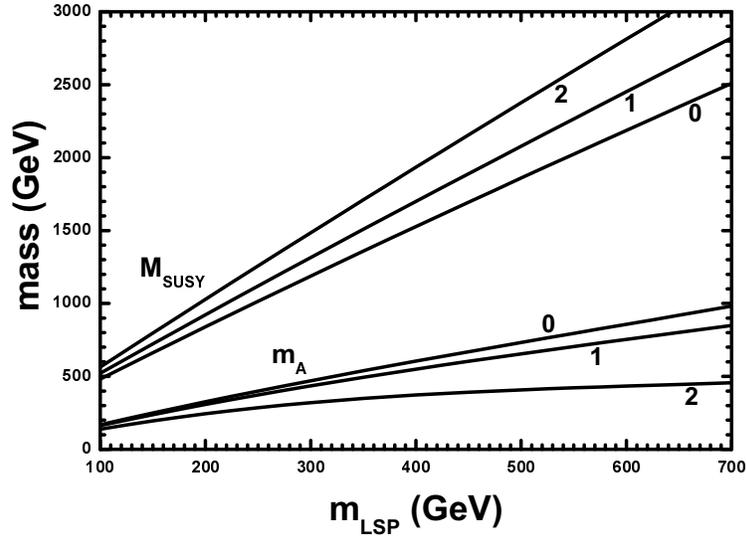,angle=-90,width=11.cm}} \hfill
\caption{\sl\ftn The mass parameters $m_A$ and $M_{\rm SUSY}$
versus $m_{\rm LSP}$ for $\mu>0$, $m_b(M_Z)=2.888~{\rm GeV}$ and
various values of $\Delta_{\tilde\tau_2}$, which are indicated on
the curves.}\label{figb}
\end{figure}

\begin{figure}[!ht]
\centerline{\epsfig{file=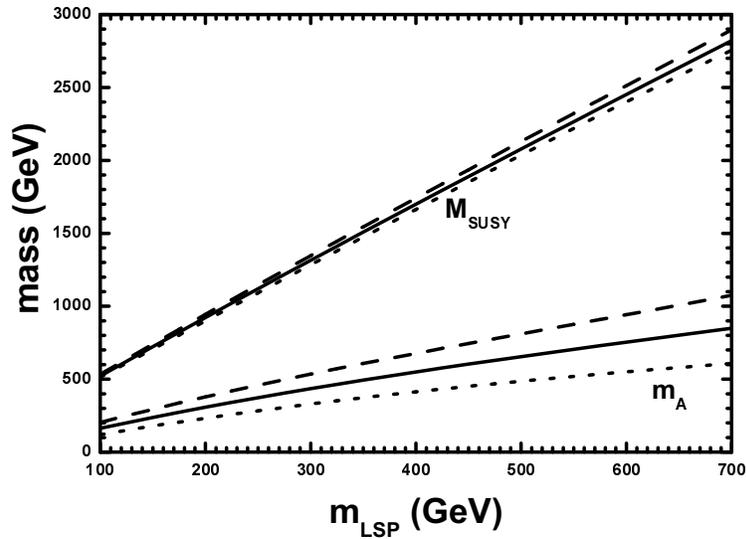,angle=-90,width=11.cm}} \hfill
\caption{\sl\ftn The mass parameters $m_A$ and $M_{\rm SUSY}$
versus $m_{\rm LSP}$ for $\mu>0$, $\Delta_{\tilde\tau_2}=1$ and
with $m_b(M_Z)=2.684~{\rm GeV}$ (dashed lines), $3.092~{\rm GeV}$
(dotted lines) or $2.888~{\rm GeV}$ (solid lines).} \label{figc}
\end{figure}

We observe the following:

\begin{itemize}
\item The lower bounds on $m_{\rm LSP}$ are not so sensitive to the
variations of $m_b(M_Z)$.
\item The lower bound on $m_{\rm LSP}$ from Eq. (\ref{mhb})
overshadows all others.
\item The LSP annihilation via the $s$-channel exchange of
an $A$-boson is by far the dominant (co)annihilation process near
the almost vertical part of the line corresponding to the upper
bound on $m_{\rm LSP}$ from Eq. (\ref{cdmb}). This can be
explained by considering Fig. \ref{figb}, where we draw $m_A$ and
$M_{\rm SUSY}$ versus $m_{\rm LSP}$ for various
$\Delta_{\tilde\tau_2}$'s and the central value of $m_b(M_Z)$. We
see that $m_A$ is always smaller than $2m_{\rm LSP}$ but close to
it. We also observe that, as $m_{\rm LSP}$ or
$\Delta_{\tilde\tau_2}$ increases, we move away from the $A$-pole
which, thus, becomes less efficient. As a consequence,
$\Omega_{\rm LSP}h^2$ increases with $m_{\rm LSP}$ or
$\Delta_{\tilde\tau_2}$.

\item The upper bound on $m_{\rm LSP}$ from Eq. (\ref{cdmb}) is
extremely sensitive to the variations of $m_b(M_Z)$. Especially
sensitive is the almost vertical part of the line corresponding to
this bound, where, as we saw above, the LSP annihilation via an
$A$-boson exchange in the $s$-channel is by far the dominant
process. This extreme sensitivity can be understood from Fig.
\ref{figc}, where $m_A$ is depicted versus $m_{\rm LSP}$ for the
lower, upper and central values of $m_b(M_Z)$ in Eq. (\ref{mbrg}).
We see that, as $m_b(M_Z)$ decreases, $m_A$ increases and
approaches $2m_{\rm LSP}$. The $A$-pole annihilation is then
enhanced and $\Omega_{\rm LSP}h^2$ is drastically reduced causing
an increase of the upper bound on $m_{\rm LSP}$.

\item For $\Delta_{\tilde\tau_2}<0.25$, bino-stau
coannihilations \cite{ellis} take over leading to a very
pronounced reduction of $\Omega_{\rm LSP}h^2$, thereby increasing
the upper limit on $m_{\rm LSP}$.

\end{itemize}

For $m_b(M_Z)=2.888~{\rm GeV}$, the allowed ranges of $m_{\rm
LSP}$ and $\Delta_{\tilde\tau_2}$ are
\beq 176~{\rm GeV}\lesssim m_{\rm LSP}\lesssim615~{\rm
GeV}~~\mbox{and}~~0\lesssim\Delta_{\tilde\tau_2}\lesssim1.8.
\label{mLSPp}\eeq

%%%%%%%%%%%%%%%%%%%%%%%%%%%%%%%%%%%%%%%%%%%%%%%%%%%%%%%%%%%%%%%%%%%%
%%%%%%%%%%%%%%%%%%%%%%%%%%%%%%%%%%%%%%%%%

%%%%%%%%%%%%%%%%%%%%%%%%%%%%%%%%%%%%%%%%%%%%%%%%%%%%%%%%%%%%%%%%%%%%
%%%%%%%%%%%%%%%%%%%%%%%%%%%%%%%%%%%%%%%%%

\subsection{The {\boldmath $\mu<0$} Case}\label{rmuneg}

As is seen from Fig. \ref{Mnx}, $2m_{\rm LSP} \gg m_A$ in the
$\mu<0$ case. So, the LSP annihilation to $b\bar b$ via the
$s$-channel exchange of an $A$-boson is not enhanced as for
$\mu>0$ (the important annihilation channels, for $\mu<0$, are not
only the ones with fermions $f\bar f$ in the final state, but also
with $HZ$, $W^\pm H^\mp$ and $hA$ \cite{pallis,nra}). As a
consequence, the only available mechanism for reducing the
$\Omega_{\rm LSP}h^2$ is the $\tilde\chi-\tilde\tau_2$
coannihilation \cite{ellis} (for an updated study, see also Ref.
\cite{roberto}), which becomes efficient when
$\Delta_{\tilde\tau_2}<0.25$.

%%%%%%%%%%%%%%%%%%%%%%%%%%%%%%%%%%%%%%%%%%%%%%%%%%%%%%%%%%%%%%%%%%%%
\begin{figure}[t]
\centerline{\epsfig{file=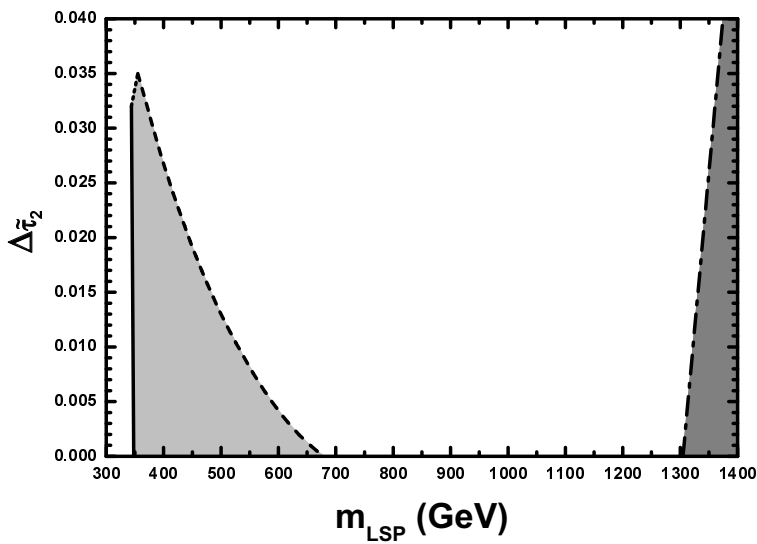,angle=-90,width=11.cm}} \hfill
\caption{\sl\ftn Restrictions on the $m_{\rm
LSP}-\Delta_{\tilde\tau_2}$ plane for $\mu<0$ and $m_b(M_Z)$ in
the range of Eq. (\ref{mbrg}). The solid [dot-dashed] line
corresponds to the lower bound on $m_{\rm LSP}$ from Eq.
(\ref{g2t}{\sf a})~[Eq. (\ref{bsgb}{\sf b})]. The dashed [dotted]
line corresponds to the upper bound on $m_{\rm
LSP}~[\Delta_{\tilde\tau_2}]$ from Eq. (\ref{cdmb}).}\label{A0x}
\end{figure}
%%%%%%%%%%%%%%%%%%%%%%%%%%%%%%%%%%%%%%%%%

\par
The restrictions from all the  requirements imposed on the $m_{\rm
LSP}-\Delta_{\tilde\tau_2}$ plane for any $m_b(M_Z)$ in Eq.
(\ref{mbrg}) are presented in Fig. \ref{A0x}. The lower bound on
$m_{\rm LSP}$ from Eq. (\ref{g2t}{\sf a}) corresponds to
$m_b(M_Z)\simeq 3.092~{\rm GeV}$ and is represented by a solid
line. The maximal $\Delta_{\tilde\tau_2}(\simeq0.032)$ on this
line yields $\Omega_{\rm LSP}h^2\simeq0.13$ for the same value of
$m_b(M_Z)$. As $m_b(M_Z)$ decreases, the maximal
$\Delta_{\tilde\tau_2}$ from Eq. (\ref{cdmb}) increases along the
dotted line and reaches its overall maximal value at
$\Delta_{\tilde\tau_2}\simeq0.035$ for $m_b(M_Z)\simeq 2.684~{\rm
GeV}$. The upper bound on $m_{\rm LSP}$ from Eq. (\ref{cdmb}) is
achieved for $m_b(M_Z)\simeq 2.684~{\rm GeV}$ and corresponds to
the dashed line. In the lightly shaded region allowed by Eqs.
(\ref{g2t}{\sf a}) and (\ref{cdmb}), we find
\beq348~{\rm GeV}\lesssim m_{\rm LSP}\lesssim 680~{\rm
GeV}~~\mbox{and}~~0\lesssim\Delta_{\tilde\tau_2}\lesssim0.035.
\label{ana}\eeq
The maximal $m_{\rm LSP}$ is achieved at $\Delta_{\tilde\tau_2}=0$
yielding ${\rm BR} (b\rightarrow s\gamma)\simeq5.8 \times10^{-4}$.

On the other hand, the lower bound on $m_{\rm LSP}$ (dot-dashed
line) from Eq. (\ref{bsgb}{\sf b}) corresponds to
$m_b(M_Z)\simeq3.092~{\rm GeV}$. In the corresponding allowed
(dark shaded) area in Fig. \ref{A0x},
\beq m_{\rm LSP} \gtrsim1305.04~{\rm GeV}\label{anb} \eeq
with the minimal $m_{\rm LSP}$ achieved at
$\Delta_{\tilde\tau_2}=0$ and yielding $\Omega_{\rm
LSP}h^2\simeq0.65$. Note that the NLO corrections to ${\rm
BR}(b\to s\gamma)$ (see Sec. \ref{phenog}) drastically reduce this
lower bound on $m_{\rm LSP}$. It is worth mentioning that, in the
second paper of Ref.~\cite{cdm} (which adopts the opposite sign
convention for $\mu$), where the $\tan\beta$ enhanced and NLO SUSY
QCD corrections were not included, the reduction was considerably
higher and yielded a much less stringent restriction.

Combining Eqs. (\ref{ana}) and (\ref{anb}), it is obvious that we
are left with no simultaneously allowed region. Needless to say
that the constraints from Eqs. (\ref{bsgb}{\sf a}) and
(\ref{g2t}{\sf b}) do not restrict the parameters, since they are
always satisfied for $\mu<0$. The same is also valid for the bound
on the lightest Higgs boson mass, Eq. (\ref{mhb}), due to the
heavy SUSY spectrum.

\section{The Deviation from YU}\label{delta}

The deviation from YU is estimated by employing Eq. (\ref{dhdef}).
In the allowed (shaded) area of Fig. \ref{figa} (for $\mu>0$)
which corresponds to the central value of $m_b(M_Z)$ in
Eq.~(\ref{mbrg}), the ranges of the parameters $c$, $\delta
h_{\tau}, \delta h_b$  and $\tan\beta$ are
\bea 0.14\lesssim c\lesssim0.17,\>0.24\lesssim\delta
h_{\tau}=-\delta h_b \lesssim0.29\>\>\mbox{and}\>\>58\lesssim
\tan\beta\lesssim 59. \nonumber \eea
However, letting $m_b(M_Z)$ vary in its 95$\%$ c.l. range in
Eq.~(\ref{mbrg}), we find that, in the corresponding allowed area,
these parameters range as follows:
\bea 0.11\lesssim c\lesssim0.19,\>0.2\lesssim\delta
h_{\tau}=-\delta h_b \lesssim0.32\>\>\mbox{and}\>\>57 \lesssim
\tan\beta\lesssim 60. \nonumber \eea

In the lightly shaded area of Fig. \ref{A0x} (for $\mu<0$), which
is obtained by allowing $m_b(M_Z)$ to vary in the range of Eq.
(\ref{mbrg}), the parameters above range as follows:
\bea 0.01\lesssim -c\lesssim 0.06,\>0.04\lesssim-\delta
h_{\tau}\lesssim0.23,\>\delta
h_b=0\>\>\mbox{and}\>\>46\lesssim\tan\beta\lesssim49. \nonumber
\eea
For $m_b(M_Z)\simeq 2.888~{\rm GeV}$, in the corresponding allowed
area, these ranges become
\bea 0.036\lesssim -c\lesssim 0.045,\>0.14\lesssim-\delta
h_{\tau}\lesssim0.17,\>\delta
h_b=0\>\>\mbox{and}\>\>47\lesssim\tan\beta\lesssim48. \nonumber
\eea

We observe that, as we increase [decrease] $m_b(M_Z)$ for $\mu>0$
[$\mu<0$], the parameter $|c|$ decreases and we get closer to
exact YU. This behavior is certainly consistent with the fact (see
also Sec. \ref{sec:intro}) that the value of $m_b(M_Z)$ which
corresponds to exact YU lies well above [a little below] its
$95\%$ c.l. range for $\mu>0~[\mu<0]$.

Note, finally, that, for $\mu>0$, the required deviation from YU
is not so small. In spite of this, the restrictions from YU are
not completely lost but only somewhat weakened. In particular,
$\tan\beta$ remains large and close to 60. Actually, our model is
much closer to YU than generic models where the Yukawa coupling
constants can differ even by orders of magnitude. Also, the
deviation from YU is generated here in a natural, systematic,
controlled and well-motivated way.

\section{Direct Detection of Neutralinos}\label{phenoaa}

As we showed in Sec. \ref{rpos}, our $\mu>0$ model possesses a
wide and natural range of parameters allowed by all the relevant
astrophysical and experimental constraints. It would be, thus,
interesting to investigate whether the predicted LSPs in the
universe could be detected in the current or planned experiments.
This can be done by first calculating the elastic scattering of
the LSPs with nuclei \cite{munoz, early}. To accomplish this goal,
we need an effective Lagrangian (see Sec. \ref{eff}) derived from
the MSSM Lagrangian (see Sec. \ref{lang}) and providing a reliable
description of the neutralino$-$quark interaction. We also need a
procedure for going from the quark to the nucleon level  (see
Sec.~\ref{phenoaa1}) and from the nucleon to the nucleus (see
Secs.~\ref{phenoaa2}, \ref{si} and \ref{sd}).

\subsection{Scalar Neutralino{\boldmath $-$}Proton Cross Section}
\label{phenoaa1}

The quantity which is being conventionally used in the recent
literature (see e.g. Refs.~\cite{kov, efo, gomez}) for comparing
experimental \cite{exp, expp, expd} and theoretical results is the
spin independent (SI) neutralino$-$proton ($\tilde\chi-p$) cross
section (at zero momentum transfer) $\sigma_{\tilde\chi p}^{\rm
SI}$ calculated by applying Eq. (\ref{sSI}) with $A_{\rm N}=Z_{\rm
N}=1$. The SI effective $\tilde\chi-p$ coupling $f_p$ is
calculated using the full one-loop treatment of
Refs.~\cite{drees1, jungman} (already used in Ref.~\cite{su5b}),
which, however, agrees with the tree-level approximation (see
Eq.~(\ref{fpn})) for the values of the SUSY parameters encountered
in our model. The values in Eq. (\ref{rgis}) \cite{efo} were
adopted for the renormalization-invariant functions $f^p_{_{{\rm
T}_q}}$ (with $q=u,d,s$) needed for the calculation of $f_p$.

Combining the sensitivities of the recent \cite{exp} and planned
\cite{expp} experiments, we obtain the following observationally
interesting region:
\beq \mbox{\sf a)}~~3\times 10^{-9}~{\rm pb}\lesssim
\sigma_{\tilde\chi p}^{\rm SI}~\quad\mbox{and}~\quad
\mbox{\sf b)}~~\sigma_{\tilde\chi p}^{\rm SI}\lesssim 2 \times
10^{-6}~{\rm pb}, \label{sgmb} \eeq
for $150~{\rm GeV}\lesssim m_{\rm LSP}\lesssim 500~{\rm GeV}$. For
the allowed values of the SUSY parameters of our model,
$\sigma_{\tilde\chi p}^{\rm SI}$ lies beyond the preferred range
of DAMA \cite{expd}, $(1-10)\times 10^{-6}~\rm{pb}$, which though
has mostly been excluded by other collaborations (e.g. EDELWEISS
and ZEPLIN I \cite{exp}).

%%%%%%%%%%%%%%%%%%%%%%%%%%%%%%%%%%%%%%%%%sigmqa.eps
\begin{figure}[t]
\centerline{\epsfig{file=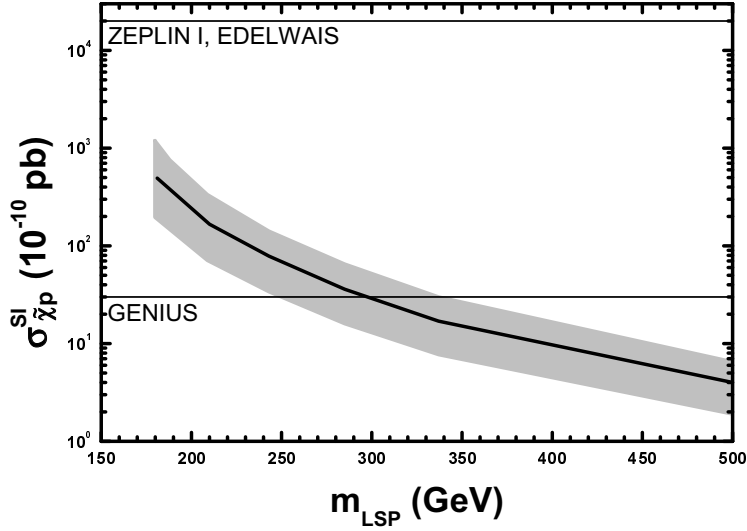,angle=-90,width=11.cm}} \hfill
\caption{\sl\ftn The SI $\tilde\chi-p$ cross section
$\sigma_{\tilde\chi p}^{\rm SI}$ versus $m_{\rm LSP}$ for $\mu>0$
and $m_b(M_Z)=2.888~{\rm GeV}$. The bold solid line is derived by
fixing $\Omega_{\rm LSP}h^2$ and $f_{_{{\rm T}_q}}^{p}$ to their
central values in Eqs. (\ref{cdmba}) and (\ref{rgis})
respectively, whereas the shaded band by allowing $\Omega_{\rm
LSP}h^2$ and $f_{_{{\rm T}_q}}^{p}$ to vary in their ranges in
these equations. The region of Eq. (\ref{sgmb}), which is
preferred by the various experimental projects, is approximately
limited between the two thin solid lines.} \label{sigmq}
\end{figure}
%%%%%%%%%%%%%fulfilling all the requirements of Sec. \ref{sec:pheno},

Allowing $\Omega_{\rm LSP}h^2$ and the hadronic inputs $f_{_{{\rm
T}_q}}^{p}$ to vary within their ranges in Eqs.~(\ref{cdmba}) and
(\ref{rgis}) respectively, we derive the shaded band on the
$m_{\rm LSP}-\sigma_{\tilde\chi p}^{\rm SI}$ plane (Fig.
\ref{sigmq}). The bold solid line corresponds to the central
values of $\Omega_{\rm LSP}h^2$ and $f_{_{{\rm T}_q}}^{p}$. Note
that the width of the band is almost exclusively due to the
variation of $f_{_{{\rm T}_q}}^{p}$ since, for fixed $m_{\rm
LSP}$, $\sigma_{\tilde\chi p}^{\rm SI}$ is almost insensitive to
the variation of $\Omega_{\rm LSP}h^2$ within the range of Eq.
(\ref{cdmba}) (or, equivalently, to the required variation of
$\Delta_{\tilde\tau_2}$). We observe that, for the allowed $m_{\rm
LSP}$'s, there are $\sigma_{\tilde\chi p}^{\rm SI}$'s which lie
within the margin of Eq. (\ref{sgmb}). Therefore, the LSPs
predicted by our model can be detectable in the near future
experiments. In particular, employing central values for
$\Omega_{\rm LSP}h^2$ and $f_{_{{\rm T}_q}}^{p}$, we find
\beq 5\times10^{-7}~{\rm pb} \gtrsim\sigma_{\tilde\chi p}^{\rm
SI}\gtrsim3\times10^{-9}~{\rm pb}~~\mbox{for}~~179.7~{\rm
GeV}\lesssim m_{\rm LSP}\lesssim300~{\rm GeV}. \label{sgmp}\eeq
These values are somewhat higher than the ones obtained in similar
estimations (see e.g. the third paper in Ref. \cite{spanos}) with
lower $\tan\beta$'s. Furthermore, the upper bound on $m_{\rm LSP}$
in Eq. (\ref{mLSPp}) implies a lower bound on $\sigma_{\tilde\chi
p}^{\rm SI}$. Namely,
\beq \sigma_{\tilde\chi p}^{\rm SI} \gtrsim
1.6~(0.8)\times10^{-10}~{\rm pb},\label{sgmxp}\eeq
where the bound in parenthesis is derived by allowing $f_{_{{\rm
T}_q}}^{p}$ to vary.

\subsection{Detection Rate of Neutralinos}
\label{phenoaa2}

%%%%%%%%%%%%%%%%%%%%%%%%%%%%%%%%%%%%%%%%%%%%%%%%%%%%%%%%%%%%%%%%%%%%
\begin{figure}[t]
\centerline{\epsfig{file=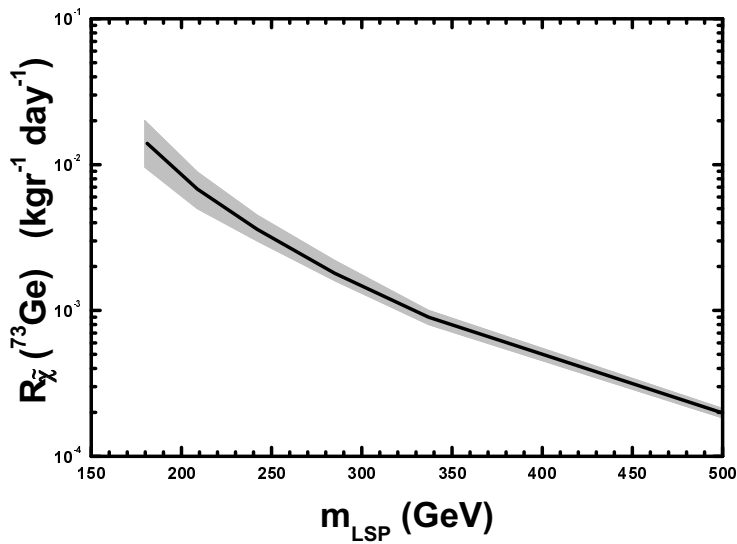,angle=-90,width=11.cm}} \hfill
\caption{\sl\ftn The LSP detection rate $R_{\tilde\chi}(^{73}{\rm
Ge})$ for a $^{73}{\rm Ge}$ detector versus $m_{\rm LSP}$ for
$\mu>0$ and $m_b(M_Z)=2.888~{\rm GeV}$. The solid line is derived
by fixing $\Omega_{\rm LSP}h^2$, $f_{_{{\rm T}_q}}^{p[n]}$ and
$\Delta^{p[n]}_q$ to their central values in Eqs. (\ref{cdmba}),
(\ref{rgis}) [(\ref{fTn})] and (\ref{Dqp}) [(\ref{Dqn})]
respectively, whereas the shaded band by allowing $\Omega_{\rm
LSP}h^2$, $f_{_{{\rm T}_q}}^{p[n]}$ and $\Delta^{p[n]}_q$ to vary
in their ranges in these equations.} \label{ratex}
\end{figure}
%%%%%%%%%%%%%%%%%%%%%%%%%%%%%%%%%%%%%%%%%
The total detection rate (events per day) of LSPs per kgr of
detector material (consisting of nuclei N) can be found by
\cite{jungman, baera}
\beq R_{\tilde\chi}({\rm
N})=\frac{\rho^0_{\tilde\chi}}{\sqrt{\pi}v_0m_{\tilde\chi}\mu_{\tilde\chi{\rm
N}}}\int_{Q_{\rm T}}^{\infty}dQ_{\rm r}\left(\sigma^{\rm
SI}_{\tilde\chi{\rm N}} F^2(Q_{\rm r})+ \sigma^{\rm
SD}_{\tilde\chi{\rm N}}\frac{S(Q_{\rm r})}{S(0)}\right)T(Q_{\rm
r})\,.\label{rate}\eeq
Here $Q_{\rm r}$ is the (recoil) energy transferred to the
nucleus, $Q_{\rm T}$ is the detector threshold energy below which
the detector is insensitive to LSP$-$nucleus recoils,
$\mu_{\tilde\chi{\rm N}}$ is the reduced mass (evaluated in Eq.
(\ref{sSI}), $\sigma^{\rm SI~[SD]}_{\tilde\chi{\rm N}}$ is the SI
[spin dependent (SD)] $\tilde\chi-{\rm N}$ cross section
(evaluated in Eq. (\ref{sSI}) [Eq.~(\ref{sSD})]) at zero $Q_{\rm
r}$, $\rho^0_{\tilde\chi}$ is the total local CDM density and
$v_0$ is the circular velocity of the sun w.r.t. the galactic rest
frame. Given the significant uncertainties involved in the
determination of the two last quantities, we adopt their most
popular values $0.3~{\rm GeV/cm^3}$ and $220~{\rm km/sec}$
respectively. The derivation of the SI and SD form factors
$F(Q_{\rm r})$ and $S(Q_{\rm r})/S(0)$ is also subject to
considerable uncertainty. They are determined by closely following
the analysis of Ref. \cite{baera} (and Ref. \cite{jungman}).
There, we can also find the function $T(Q_{\rm r})$ which depends
on the minimal velocity of the incident LSPs and the velocity of
earth (w.r.t. the galactic rest frame), which generates an annual
modulation \cite{vergados}.\vskip0.19cm

In Fig. \ref{ratex}, we display $R_{\tilde\chi}(^{73}{\rm Ge})$ as
a function $m_{\rm LSP}$ for $Q_{\rm T}=11~{\rm keV}$ and for the
second of June. The shaded region is derived by allowing
$\Omega_{\rm LSP}h^2$, $f_{_{{\rm T}_q}}^{p[n]}$ and
$\Delta^{p[n]}_q$ (see Secs.~\ref{si} and \ref{sd}) to vary within
their ranges in Eqs.~(\ref{cdmba}), (\ref{rgis}) [(\ref{fTn})] and
(\ref{Dqp}) [(\ref{Dqn})] respectively, whereas the solid line by
fixing them to their central values. Note that the width of the
band is almost exclusively due to the variation of $\sigma^{\rm
SI}_{\tilde\chi{\rm N}}$, since $\sigma^{\rm SD}_{\tilde\chi{\rm
N}}$ is almost insensitive to the aforementioned variations and an
order of magnitude smaller than $\sigma^{\rm SI}_{\tilde\chi{\rm
N}}$. The resulting $R_{\tilde\chi}(^{73}{\rm Ge})$ is a little
lower than the one obtained with the naive choice
$F(Q_r)=S(Q_r)/S(0)=1$ and significantly lower than the one
corresponding to a wino or Higgsino LSP (compare with Ref.
\cite{murakami}).

\section{A Resolution of the {\boldmath $\mu$} Problem}
\label{sec:rsym}

An important shortcoming of MSSM is that there is no understanding
of how the SUSY $\mu$ term, with the right magnitude of $|\mu|\sim
10^{2}-10^{3}~{\rm GeV}$, arises. One way \cite{rsym} to solve
this $\mu$ problem is via a PQ symmetry $U(1)_{\rm PQ}$ \cite{pq},
which also solves the strong CP problem. This solution is based on
the observation \cite{kn} that the axion decay constant $f_{a}$,
which is the symmetry breaking scale of $U(1)_{\rm PQ}$, is
(normally) intermediate ($\sim 10^{11}-10^{12}~{\rm GeV}$) and,
thus, $|\mu|\sim f_{a}^2/M_{\rm S}$. The scale $f_{a}$ is, in
turn, $\sim (m_{3/2}M_{\rm S})^{1/2}$, where $m_{3/2}\sim
1~{\rm{TeV}}$ is the gravity-mediated soft SUSY breaking scale
(gravitino mass). In order to implement this solution of the $\mu$
problem, we introduce \cite{rsym} a pair of superfields $N$ and
$\bar{N}$ (see Table 1) with the following non-renormalizable
couplings in the superpotential \cite{laz4}:
\beq W_{\rm PQ} = \lambda_{\mu} \frac{N^2 h^2}{M_{\rm S}}
+\lambda^\prime_{\mu} \frac{N^2 h^{\prime2}}{M_{\rm S}} +\lambda
\frac{N^2 H^c\bar H^c h^\prime h}{M^3_{\rm S}} + \lambda_{\rm PQ}
\frac{N^2 \bar{N}^2}{M_{\rm S}}\cdot\label{eq:superpot} \eeq
Here, $\lambda_\mu$, $\lambda^\prime_\mu$, $\lambda$ and
$\lambda_{\rm PQ}$ are taken positive by redefining the phases of
$N$ and $\bar{N}$. After SUSY breaking, the $N^2\bar N^2$ term
leads to the scalar potential
\begin{eqnarray}
V_{\rm PQ}&=&\left(m_{3/2}^2 +4\lambda_{\rm
PQ}^2\left|\frac{N\bar{N}}{M_{\rm S}}\right|^2\right)
\left[(|N|-|\bar{N}|)^2+2|N||\bar{N}|\right] \nonumber \\ &
&+2|A|m_{3/2}\lambda_{\rm PQ}\frac{|N\bar{N}|^2}{M_{\rm S}}
{\rm{cos}}(\epsilon+2\theta+2\bar{\theta}), \label{eq:pqpot}
\end{eqnarray}
where $A$ is the dimensionless coefficient of the soft SUSY
breaking term corresponding to the superpotential term
$N^2\bar{N}^2$ and $\epsilon$, $\theta$, $\bar{\theta}$ are the
phases of $A$, $N$, $\bar{N}$ respectively. Minimization of
$V_{\rm PQ}$ then requires $|N|=|\bar{N}|$,
$\epsilon+2\theta+2\bar{\theta}=\pi$ and $V_{\rm PQ}$ takes the
form
\begin{equation}
V_{\rm PQ}=2|N|^2m_{3/2}^2\left(4\lambda_{\rm PQ}^2\frac{|N|^4}
{m_{3/2}^2M_{\rm S}^2}-|A|\lambda_{\rm
PQ}\frac{|N|^2}{m_{3/2}M_{\rm S}} +1\right). \label{eq:pqpotmin}
\end{equation}
For $|A|>4$, the absolute minimum of the potential is at
\begin{equation}
|\langle N\rangle|=|\langle\bar{N}\rangle|\equiv
\frac{f_a}{2}=\sqrt{m_{3/2}M_{\rm S}}\
\sqrt{\frac{|A|+\sqrt{|A|^2-12}}{12\lambda_{\rm PQ}}}\sim
\sqrt{m_{3/2}M_{\rm S}}. \label{eq:solution}
\end{equation}
The $\mu$ term is generated predominantly via the terms $N^2h^2$
and $N^2h^{\prime2}$ of Eq. (\ref{eq:superpot}) with
$|\mu|\sim|\langle N\rangle|^2/M_{\rm S}$, which is of the right
magnitude.

The potential $V_{\rm PQ}$ also has a local minimum at
$N=\bar{N}=0$, which is separated from the global PQ minimum by a
sizable potential barrier preventing a successful transition from
the trivial to the PQ vacuum. This situation persists at all
cosmic temperatures after reheating, as has been shown \cite{jean}
by considering the one-loop temperature corrections \cite{jackiw}
to the scalar potential. We are, thus, obliged to assume that,
after the termination of inflation, the system emerges with the
appropriate combination of initial conditions so that it is led
\cite{curvaton} to the PQ vacuum.

\section{Proton Stability} \label{sec:prot}

We can assign baryon number $1/3~[-1/3]$ to all color triplets
[antitriplets]. Recall that there are (anti)triplets not only in
$F, F^c$ but also in $H^c, \bar{H}^c, G$. Lepton number is then
defined via $B-L$. Before the inclusion of the extra Higgs
superfields (see Table 1), $B$ (and $L$) violation comes from the
following terms \cite{jean}
\begin{equation}
F^c F^c H^c H^c,~~ F F \bar{H}^c \bar{H}^c h h,~~F F \bar{H}^c
\bar{H}^c \bar{N}^2  \label{terms}
\end{equation}
(and terms containing the combinations $(H^c)^4$,
$(\bar{H}^c)^4$), which give couplings like $u^c d^c d^c_H
\nu^c_H$ (or $u^c d^c u^c_H e^c_H$),\ $u d \bar{d}^c_H
\bar{\nu}^c_H$ (or $u d \bar{u}^c_H \bar{e}^c_H$) with appropriate
coefficients. Also, the terms $G H^c H^c$ and $G \bar{H}^c
\bar{H}^c$ give rise to the $B$ (and $L$) violating couplings $g^c
u^c_H d^c_H$,\ $\bar{g}^c \bar{u}^c_H \bar{d}^c_H$. All other
combinations are $B$ (and $L$) conserving since all their
$SU(4)_{\rm c}$ {\bf 4}'s are contracted with ${\bf \bar4}$'s.

The dominant contribution to proton decay comes from effective
dimension five operators generated by one-loop diagrams with two
of the $u^c_H$, $d^c_H$ or one of the $u^c_H$, $d^c_H$ and one of
the $\nu^c_H$, $e^c_H$ circulating in the loop. The amplitudes
corresponding to these operators are estimated to be at most of
order $m_{3/2}M_{\rm GUT}/M_{\rm S}^3 \lesssim 10^{-34}~
{\rm{GeV}}^{-1}$. This makes the proton practically stable.

After the inclusion of the superfields $h^{\prime}$ and
$\bar{h}^{\prime}$, the couplings
\beq FF\bar{H}^c\bar{H}^chh^{\prime},
~FF\bar{H}^c\bar{H}^ch^{\prime}h^{\prime} \label{couplings} \eeq
(as well as the new couplings containing arbitrary powers of the
combinations $(H^c)^4$, $(\bar{H}^c)^4$) give rise \cite{qcdm} to
additional $B$ and $L$ number violation. However, their
contribution to proton decay is subdominant to the one arising
from the terms of Eq. (\ref{terms}). One can further show
\cite{qcdm} that the inclusion of the superfields $\phi$ and
$\bar\phi$ also gives a subdominant contribution to the proton
decay.

\section{The Inflationary Scenario} \label{inflation}

One of the most promising inflationary scenarios is hybrid
inflation \cite{linde}, which uses two real scalars: one which
provides the vacuum energy density for inflation and a second
which is the slowly varying field during inflation. This scheme is
naturally incorporated \cite{hybrid} in SUSY GUTs (for an updated
review, see Ref. \cite{senoguz}), but in its standard realization
has the following problem \cite{pana1}: if the GUT gauge symmetry
breaking predicts monopoles (and this is the case of $G_{\rm PS}$
which predicts doubly charged monopoles \cite{laz3}), they are
copiously produced at the end of inflation leading to a
cosmological catastrophe \cite{kibble}. One way to remedy this is
to generate a shifted inflationary trajectory so that $G_{\rm PS}$
is already broken during inflation. This could be achieved
\cite{jean} in our SUSY GUT model even before the introduction of
the extra Higgs superfields (see Table 1), but only by utilizing
non-renormalizable terms. The inclusion of $h^\prime$ and $\bar
h^\prime$ does not change this situation, which thus also holds in
our $\mu<0$ model. This model is though excluded by imposing the
restrictions of Sec.~\ref{sec:pheno}. On the other hand, for
$\mu>0$, the inclusion of $\phi$ and $\bar\phi$ very naturally
gives rise \cite{jean2} to a shifted path, but only with
renormalizable interactions (for similar recent analyses in the
context of a $SU(5)$ SUSY GUT, see Ref. \cite{kye}).

\subsection{The Shifted Inflationary Path} \label{inflationw}

As we showed in Sec. \ref{mupos}, the deviation from YU which is
needed for $\mu>0$ can be achieved by the inclusion of the
superfields $\phi$, $\bar{\phi}$. In the presence of these
superfields, a new version \cite{jean2} of shifted hybrid
inflation \cite{jean} can take place, without invoking any
non-renormalizable superpotential terms.

Indeed, these fields lead to three new renormalizable terms in the
part of the superpotential which is relevant for inflation. This
is given by
\begin{equation}
W=\kappa S(H^c\bar{H}^c-M^2)-\beta S\phi^2+
m\phi\bar{\phi}+\lambda\bar{\phi}H^c\bar{H}^c,
\label{superpotential}
\end{equation}
where $M,m \sim M_{\rm GUT}$, and $\kappa$, $\beta$ and $\lambda$
are dimensionless coupling constants with
$M,~m,~\kappa,~\lambda>0$ by field redefinitions. For simplicity,
we take $\beta>0$ (the parameters are normalized so that they
correspond to the couplings between the SM singlet components of
the superfields).

\par
The scalar potential obtained from $W$ is given by
\begin{eqnarray}
V=\left\vert\kappa(H^c\bar{H}^c-M^2)-\beta\phi^2
\right\vert^2+\left\vert 2\beta S\phi-m\bar{\phi}
\right\vert^2+\left\vert m\phi+\lambda H^c
\bar{H}^c\right\vert^2\nonumber \\ +\left\vert\kappa
S+\lambda\bar{\phi} \right\vert^2\left(\vert
H^c\vert^2+\vert\bar{H}^c \vert^2\right)+{\rm
D-terms}.~~~~~~~~~~~~~~~~ \label{potential}
\end{eqnarray}
Vanishing of the D-terms yields
$\bar{H}^c\,^{*}=e^{i\vartheta}H^c$ ($H^c$, $\bar{H}^c$ lie in the
$\nu^c_H$, $\bar{\nu}^c_H$ direction). We restrict ourselves to
the direction with $\vartheta=0$ which contains the shifted
inflationary path and the SUSY vacua (see below). Performing
appropriate R and gauge transformations, we bring $H^c$,
$\bar{H}^c$ and $S$ to the positive real axis.

\par
From the potential in Eq. (\ref{potential}), we find that the SUSY
vacuum lies at
\begin{equation}
\frac{H^c\bar{H}^c}{M^2}\equiv\left(\frac{v_0}
{M}\right)^2=\frac{1}{2\xi}\left(1-\sqrt{1-4\xi}\right),
~~\frac{\phi}{M}=-\sqrt{\frac{\kappa\xi}{\beta}}\left(
\frac{v_0}{M}\right)^2 \label{vacuum}
\end{equation}
with $S=0$ and $\bar{\phi}=0$, where $\xi=\beta\lambda^2M^2/\kappa
m^2<1/4$. The potential possesses a `shifted' flat direction
(besides the trivial one) at
\begin{equation}
\frac{H^c\bar{H}^c}{M^2}\equiv\left(\frac{v}{M}
\right)^2=\frac{2\kappa^2(\frac{1}{4\xi}+1)+
\frac{\lambda^2}{\xi}}{2(\kappa^2+\lambda^2)},
~~\frac{\phi}{M}=-\frac{1}{2}\sqrt{\frac{\kappa}{\beta\xi}},
~~\bar{\phi}=-\frac{\kappa}{\lambda}S \label{trajectory}
\end{equation}
with $S>0$ and a constant potential energy density $V_0$ given by
\begin{equation}
\frac{V_0}{M^4}=\frac{\kappa^2\lambda^2}{\kappa^2+
\lambda^2}\left(\frac{1}{4\xi}-1\right)^2, \label{V0}
\end{equation}
which can be used as inflationary path. $V_0\neq0$ breaks SUSY  on
this path, while the constant non-zero values of $H^c$,
$\bar{H}^c$ break the GUT gauge symmetry too. The SUSY breaking
implies the existence of one-loop radiative corrections
\cite{DvaSha} which lift the classical flatness of this path
yielding the necessary inclination for driving the inflaton
towards the SUSY vacuum.

\par
The one-loop radiative corrections to $V$ along the shifted
inflationary trajectory are calculated by using the
Coleman-Weinberg formula \cite{cw}:
\begin{equation}
\Delta V=\frac{1}{64\pi^2}\sum_i(-)^{F_i}M_i^4\ln
\frac{M_i^2}{\Lambda^2}, \label{Coleman}
\end{equation}
where the sum extends over all helicity states $i$, $F_i$ and
$M_i^2$ are the fermion number and mass squared of the $i$th state
and $\Lambda$ is a renormalization mass scale. In order to use
this formula for creating a logarithmic slope which drives the
canonically normalized real inflaton field
$\sigma=\sqrt{2(\kappa^2+\lambda^2)}S/\lambda$ towards the
minimum, one has first to derive the mass spectrum of the model on
the shifted inflationary path. This is a quite complicated task
and we will skip it here.

\subsection{Inflationary Observables} \label{inflationq}

The slow roll parameters are given by (see e.g. Ref.
 \cite{lectures})
\begin{equation}
\epsilon\simeq\frac{m_{\rm P}^2}{2}~\left(
\frac{V'(\sigma)}{V_0}\right)^2~~\mbox{and}~~\eta\simeq m_{\rm
P}^2~\frac{V''(\sigma)} {V_0}, \label{slowroll}
\end{equation}
where the primes denote derivation w.r.t. the real normalized
inflaton field $\sigma$ and $m_{\rm P}\simeq 2.44\times
10^{18}~{\rm GeV}$ is the reduced Planck scale. The conditions for
inflation to take place are $\epsilon\leq 1$ and
$\vert\eta\vert\leq 1$.

\par
The number of e-foldings that our present horizon scale suffered
during inflation can be calculated as follows (see e.g. Ref.
 \cite{lectures}):
\begin{equation}
N_Q\simeq\frac{1}{m_{\rm P}^2}\int_{\sigma_f}
^{\sigma_Q}\frac{V_0}{V'(\sigma)}d\sigma \simeq\ln\left(4.41\times
10^{11}~ T_r^{\frac{1}{3}}~V_0^{\frac{1}{6}} \right), \label{NQ}
\end{equation}
where $\sigma_f~[\sigma_Q]$ is the value of $\sigma$ at the end of
inflation [when our present horizon scale crossed outside the
inflationary horizon] and $T_r\simeq 10^9~{\rm GeV}$ is the reheat
temperature taken to saturate the gravitino constraint
 \cite{gravitino}.

The quadrupole anisotropy of the CMBR can be calculated as follows
(see e.g. Ref.  \cite{lectures}):
\begin{equation}
\left(\frac{\delta T}{T}\right)_Q\simeq\frac{1}
{12\sqrt{5}}\frac{\sqrt{V_0^3}}{m_{\rm P}^3V'(\sigma_Q)}\cdot
\label{anisotropy}
\end{equation}
Fixing $(\delta T/T)_Q\simeq 6.6\times 10^{-6}$ to its central
value from COBE  \cite{cobe} (under the condition that the
spectral index $n=1$), we can determine one of the free parameters
(say $\beta$) in terms of the others ($m$, $\kappa$ and
$\lambda$). For instance, one finds \cite{jean2} $\beta=0.1$ for
$m=4.35 \times 10^{15}~{\rm GeV}$ and $\kappa=\lambda= 3\times
10^{-2}$. In this case, the instability point of the shifted path
is at $\sigma_c\simeq 3.55\times 10^{16}~{\rm GeV}$,
$\sigma_f\simeq 1.7\times 10^{17}~{\rm GeV}$ and $\sigma_Q \simeq
1.6\times 10^{18}~{\rm GeV}$. Also, $M\simeq 2.66\times
10^{16}~{\rm GeV}$, $N_Q \simeq 57.7$ and the spectral index
$n\simeq 0.98$. Note that the slow roll conditions are violated
and inflation ends well before reaching the instability point at
$\sigma_c$. We see that the COBE constraint can be easily
satisfied with natural values of the parameters. Moreover,
superheavy SM non-singlets with masses $\ll M_{\rm GUT}$, which
could disturb the unification of the MSSM gauge coupling
constants, are not encountered.

\subsection{SUGRA Corrections} \label{sec:sugra}

As we emphasized, the new shifted hybrid inflation occurs at
values of $\sigma$ which are quite close to the reduced Planck
scale. Thus, one cannot ignore the SUGRA corrections to the scalar
potential.

The scalar potential in SUGRA, without the D-terms, is given by
\begin{equation}
V=e^{K/m_{\rm P}^2}\left[(F_i)^*K^{i^*j} F_j-3\frac{\vert
W\vert^2}{m_{\rm P}^2}\right], \label{sugra}
\end{equation}
where $K$ is the K\"{a}hler potential, $F_i=W_i +K_iW/m_{\rm
P}^2$, a subscript $i~[i^*]$ denotes derivation w.r.t. the complex
scalar field $\phi^i~[\phi^i\,^{*}]$ and $K^{i^*j}$ is the inverse
of the matrix $K_{ji^*}$.

\par
Consider a (complex) inflaton $\Sigma$ corresponding to a flat
direction of global SUSY with $W_{i\Sigma}=0$. We assume that the
potential on this path depends only on $|\Sigma|$, which holds in
our model due to a global symmetry. From Eq. (\ref{sugra}), we
find that the SUGRA corrections lift the flatness of the $\Sigma$
direction by generating a mass squared for $\Sigma$ (see e.g.
Ref. \cite{lyth})
\begin{equation}
m_\Sigma^2=\frac{V_0}{m_{\rm P}^2}- \frac{\vert
W_\Sigma\vert^2}{m_{\rm P}^2}+\sum_{i,j}
(W_i)^*K^{i^*j}_{\Sigma^*\Sigma}W_j+\cdots, \label{mSigma}
\end{equation}
where the right hand side (RHS) is evaluated on the flat direction
with the explicitly displayed terms taken at $\Sigma=0$. The
ellipsis represents higher order terms which are suppressed by
powers of $|\Sigma|/m_{\rm P}$. The slow roll parameter $\eta$
then becomes
\begin{equation}
\eta=1-\frac{\vert W_\Sigma\vert^2}{V_0}+ \frac{m_{\rm
P}^2}{V_0}\sum_{i,j}(W_i)^* K^{i^*j}_{\Sigma^*\Sigma}W_j+\cdots,
\label{etasugra}
\end{equation}
which, in general, could be of order unity and, thus, invalidate
\cite{hybrid, stew} inflation. This is the well-known $\eta$
problem of inflation in local SUSY. Several proposals have been
made to overcome this difficulty (for a review, see e.g. Ref.
\cite{lyth}).

In all versions of SUSY hybrid inflation, there is an automatic
mutual cancellation between the first two terms in the RHS of Eq.
(\ref{etasugra}). This is due to the fact that $W_n=0$ on the
inflationary path for all field directions $n$ which are
perpendicular to this path, which implies that $|W_\Sigma|^2=V_0$
on the path. This is an important feature since, in general, the
sum of the first two terms in the RHS of Eq. (\ref{etasugra}) is
positive and of order unity, thereby ruining inflation. Moreover,
$W_n=0$ also implies that the only contribution to the third term
in the RHS of Eq. (\ref{etasugra}) comes from the term in $K$
which is quartic in $\Sigma$. So the third term can be suppressed
by mildly tuning just one parameter \cite{inf} and inflation could
remain intact provided that the terms in the ellipsis can be
ignored.

However, in our present model, inflation takes place at values of
$|\Sigma|$ close to $m_{\rm P}$. So, the terms in the ellipsis in
the RHS of Eq. (\ref{etasugra}) cannot be ignored (in contrast to
the case of the old version \cite{jean} of shifted hybrid
inflation) and may easily invalidate inflation. We, thus, need to
invoke here a mechanism which can ensure that the SUGRA
corrections do not lift the flatness of the inflationary path to
all orders. A suitable scheme has been suggested in Ref.
\cite{panag}. It has been argued that special forms of the
K\"{a}hler potential can lead to the cancellation of the SUGRA
corrections which spoil slow roll inflation to all orders. In
particular, a specific form of $K(\Sigma)$ (used in no-scale SUGRA
models) was employed and a gauge singlet field $Z$ with a similar
$K(Z)$ was introduced. It was pointed out that, by assuming a
superheavy VEV for the $Z$ field through D-terms, an exact
cancellation of the inflaton mass on the inflationary trajectory
can be achieved.

The form of the K\"{a}hler potential for $\Sigma$ used in Ref.
\cite{panag} is given by
\begin{equation}
K(\vert\Sigma\vert^2)=-{\cal N} m_{\rm P}^2\ln\left(1-
\frac{\vert\Sigma\vert^2}{{\cal N} m_{\rm P}^2}\right),
\label{kaehler}
\end{equation}
where ${\cal N}=1$ or $2$; here we will take ${\cal N}=2$. In this
case, the kinetic term of the real normalized inflaton field
$\sigma$ (note that $|\Sigma|=\sigma/\sqrt{2}$) is
$(1/2)(\partial^2 K/\partial\Sigma\partial\Sigma^*)
\dot{\sigma}^2$, where the overdot denotes derivation w.r.t. the
cosmic time $t$ and ${\partial^2K}/{\partial\Sigma\partial
\Sigma^*}=(1-\sigma^2/2{\cal N} m_{\rm P}^2)^{-2}$.
Thus, the Lagrangian density on the shifted path is given by
\begin{equation}
{\cal L}=a^3(t) \left[\frac{1}{2}\dot{\sigma}^2\left(1-
\frac{\sigma^2}{2 {\cal N} m_{\rm P}^2}\right)^{-2}-
V(\sigma)\right], \label{lagrangian}
\end{equation}
where $a(t)$ is the scale factor of the universe.

\par
The evolution equation of $\sigma$ is found by varying this
Lagrangian w.r.t. $\sigma$:
\begin{equation}
\left[\ddot{\sigma}+3H\dot{\sigma}+\dot{\sigma}^2
\left(1-\frac{\sigma^2}{2{\cal N} m_{\rm P}^2}\right)^{-1}
\frac{\sigma}{{\cal N} m_{\rm P}^2}\right]\left(1-\frac{\sigma^2}
{2{\cal N} m_{\rm P}^2}\right)^{-2}+V'(\sigma)=0, \label{motion}
\end{equation}
where $H$ is the Hubble parameter. During inflation, the
`friction' term $3 H\dot{\sigma}$ dominates over the other two
terms in the brackets in Eq. (\ref{motion}). Thus, this equation
reduces to the modified inflationary equation
\begin{equation}
\dot{\sigma}=-\frac{V'(\sigma)}{3H}\left(1- \frac{\sigma^2}{2
{\cal N} m_{\rm P}^2}\right)^2. \label{infeq}
\end{equation}
Note that, for $\sigma\ll\sqrt{2{\cal N}}m_{\rm P}$, this equation
reduces to the standard inflationary equation.

\par
To derive the slow roll conditions, we evaluate the sum of the
first and the third term in the brackets in Eq.~(\ref{motion}) by
using Eq.~(\ref{infeq}):
\begin{eqnarray}
\ddot{\sigma}+\dot{\sigma}^2\left(1- \frac{\sigma^2}{2{\cal N}
m_{\rm P}^2}\right)^{-1} \frac{\sigma}{{\cal N} m_{\rm
P}^2}=\frac{V'(\sigma)} {3H^2}H'(\sigma)\dot{\sigma}\left(1-
\frac{\sigma^2}{2{\cal N} m_{\rm P}^2}\right)^{2} \nonumber\\
-\frac{V''(\sigma)} {3H}\dot{\sigma}\left(1-\frac{\sigma^2}{2{\cal
N}m_P^2} \right)^{2}+\frac{V'(\sigma)}{3H}\dot{\sigma}
\left(1-\frac{\sigma^2}{2{\cal N} m_{\rm P}^2}\right)
\frac{\sigma}{{\cal N} m_{\rm P}^2}\cdot~~ \label{sigmaddot}
\end{eqnarray}
Comparing the first two terms in the RHS of Eq. (\ref{sigmaddot})
with $H\dot{\sigma}$, we obtain
\begin{eqnarray}
\epsilon\simeq\frac{1}{2}m_{\rm P}^2\left(
\frac{V'(\sigma)}{V_0}\right)^2\left(1- \frac{\sigma^2}{2{\cal N}
m_{\rm P}^2}\right)^{2}\leq 1, \label{epsilon}\\
\vert\eta\vert\simeq m_{\rm P}^2\bigg|\frac{V''
(\sigma)}{V_0}\bigg|\left(1-\frac{\sigma^2} {2{\cal N} m_{\rm
P}^2}\right)^{2}\leq 1.~~~ \label{eta}
\end{eqnarray}
The third term in the RHS of Eq. (\ref{sigmaddot}), compared to
$H\dot{\sigma}$, yields $\sqrt{2}\sigma \epsilon^{1/2}/{\cal N}
m_{\rm P}\leq 1$, which is automatically satisfied provided that
Eq. (\ref{epsilon}) holds and $\sigma\leq {\cal N} m_{\rm
P}/\sqrt{2}$. The latter is true for the values of $\sigma$ which
are relevant here. We see that the slow roll parameters $\epsilon$
and $\eta$ now carry an extra factor $(1-\sigma^2/2{\cal N} m_{\rm
P}^2)^2\leq 1$. This leads, in general, to smaller $\sigma_f$'s.
However, in our case, $\sigma_f\ll\sqrt{2{\cal N}}m_{\rm P}$ (for
${\cal N}=2$) and, thus, this factor is practically equal to
unity. Consequently, its influence on $\sigma_f$ is negligible.

\par
The formulas for $N_Q$ and $(\delta T/T)_Q$ are now also modified
due to the presence of the extra factor $(1-\sigma^2/2{\cal N}
m_{\rm P}^2)^2$ in Eq. (\ref{infeq}). In particular, a factor
$(1-\sigma^2/2{\cal N} m_{\rm P}^2)^{-2}$ must be included in the
integrand in the RHS of Eq. (\ref{NQ}) and a factor
$(1-\sigma_Q^2/2{\cal N} m_{\rm P}^2)^{-4}$ in the RHS of Eq.
(\ref{anisotropy}). We find that, for the $\sigma$'s under
consideration, these modifications have only a small influence on
$\sigma_Q$ if we use the same input values for the free parameters
as in the global SUSY case. On the contrary, $(\delta T/T)_Q$
increases considerably. However, we can easily readjust the
parameters so that the COBE requirements are again met. For
instance, $(\delta T/T)_Q\simeq 6.6\times 10^{-6}$ is now obtained
\cite{jean2} with $m=3.8\times 10^{15}~{\rm GeV}$ keeping
$\kappa=\lambda =3\times 10^{-2}$, $\beta=0.1$ as in global SUSY.
In this case, $\sigma_c\simeq 2.7\times 10^{16}~{\rm GeV}$,
$\sigma_f\simeq 1.8\times 10^{17}~{\rm GeV}$ and $\sigma_Q\simeq
1.6\times 10^{18}~{\rm GeV}$. Also, $M\simeq 2.6\times
10^{16}~{\rm GeV}$, $N_Q\simeq 57.5$ and $n\simeq 0.99$.

\section{Conclusions} \label{con}

We studied the CMSSM with $A_0=0$ applying a suitable set of YQUCs
which originate from SUSY GUT models based on $G_{\rm PS}$. For
each sign of $\mu$, an appropriate YQUC was chosen from this set
so that an adequate deviation from YU which allows an acceptable
$m_b(M_Z)$ is ensured. We, also, imposed the constraints from the
CDM in the universe, $b\rightarrow s\gamma$, $\delta\alpha_\mu$
and $m_h$. We concluded that

\begin{itemize}

\item For $\mu>0$, there exists a wide and natural range of CMSSM
parameters which is consistent with all the above constraints. We
found that $\tan\beta$ ranges between about 58 and 61 and the
asymptotic splitting between the bottom (or tau) and the top
Yukawa coupling constants varies in the range $26-35\%$ for
central values of $m_b(M_Z)$ and $\alpha_s(M_Z)$.

\item For $\mu<0$, despite the fact that, considering the
$\tau$-based calculation for $\alpha^{\rm SM}_\mu$, the
$\delta\alpha_\mu$ and CDM criteria can be simultaneously
satisfied, the model is excluded since the $b\rightarrow s\gamma$
and CDM requirements remain incompatible. However, the deviation
from YU needed for correcting $m_b(M_Z)$ is much smaller than in
the previous case.

\end{itemize}

The predicted LSP mass in the $\mu>0$ model can be as low as about
$176~{\rm GeV}$ with $\sigma_{\tilde\chi p}^{\rm SI}$ in the range
of the sensitivity of the planned direct CDM detectors, though
with a low detection rate. Moreover, the $\mu>0$ model resolves
the $\mu$ problem of MSSM, predicts stable proton and gives rise
to a new version of shifted hybrid inflation. The inflationary
scenario relies on renonormalizable terms only, can be consistent
with the COBE constraint on the CMBR with natural values for the
relevant parameters and avoids overproduction of monopoles at the
end of inflation. Inflation takes place along a classically flat
direction, where $G_{\rm PS}$ is spontaneously broken to $G_{\rm
SM}$. A readjustment of the input values of the free parameters is
required after the introduction of a specific K\"{a}hler potential
and an extra gauge singlet with a superheavy VEV via D-terms,
which are to be included in order for the SUGRA corrections not to
invalidate inflation.

\section*{Acknowledgments}

We would like to thank M.E. G\'omez, R. Jeannerot and S. Khalil
for fruitful and pleasant collaborations from which parts of this
work are culled. This research was supported by European Union
under the RTN contracts HPRN-CT-2000-00148 and HPRN-CT-2000-00152.

%\newpage
\appendix
\setcounter{equation}{0}
\renewcommand{\theequation}{A.\arabic{equation}}
\renewcommand{\thesubsection}{A.\arabic{subsection}}
\section*{Appendix A\\ Neutralino{\boldmath $-$}Nucleus Elastic Cross Section}

In this Appendix, we sketch the derivation of the tree-level
approximation $\sigma^0_{\tilde\chi {\rm N}}$ to the total cross
section at zero momentum transfer for the elastic scattering
process $\tilde{\chi}\, {\rm N} \rightarrow \tilde{\chi}\, {\rm
N}$ with ${\rm N}$ being a nucleus target. This quantity can be
split into a scalar or SI $(\sigma^{\rm SI}_{\tilde\chi {\rm N}})$
and an axial-vector or SD $(\sigma^{\rm SD}_{\tilde\chi {\rm N}})$
part \cite{early, kov, efo, drees1,jungman}:
\begin{eqnarray}
\label{sigmat} \sigma^0_{\tilde\chi {\rm N}} = \sigma^{\rm
SI}_{\tilde\chi {\rm N}} + \sigma^{\rm SD}_{\tilde\chi {\rm N}}.
\end{eqnarray}
In the following, we present the ingredients needed for the
calculation of each part.

%--------------------------------
\subsection{The Relevant MSSM Lagrangian}\label{lang}
%--------------------------------

We present, in terms of mass eigenstates, all the Feynman rules
which are necessary for studying the neutralino-quark elastic
scattering process.

\subsubsection{Neutralino{\boldmath $-$}Squark{\boldmath $-$}Quark
Vertex {\rm (Ref. \cite{hg}, Fig. 24)}}

The Lagrangian which describes the neutralino$-$squark$-$quark
interaction is
\begin{eqnarray}
{\cal L}_{\tilde\chi\tilde{q}q} = \sqrt{2}\, g\, t_W \,\bar q
\sum_{i=1}^{2} \Big(\,g_{iL}^{q} P_L + g_{iR}^{q}\,P_R\, \Big)
\tilde\chi\tilde{q}_i+{\rm h.c.}\,,
\end{eqnarray}
where h.c. denotes the hermitian conjugate,
$P_{L[R]}=(1-[+]\gamma_5)/2$, $q$ is a quark field, $g$ is the
$SU(2)_L$ gauge coupling constant and $t_W=\tan\theta_W$. Here, we
considered general flavor-diagonal squark mixing and
$\tilde{q}_1,\>\tilde{q}_2$ are the squark mass eigenstates
defined as follows:
\begin{eqnarray}
\llgm{\tilde q_1\cr \tilde q_2}\rrgm &=& R_{\tilde q}^\tr
\llgm{\tilde q_L \cr \tilde q_R}\rrgm,~~\mbox{where}~~ R_{\tilde
q}=\llgm{c_{\tilde q} & -s_{\tilde q} \cr s_{\tilde q} & c_{\tilde
q} }\rrgm\,, \label{eq:rotation}
\end{eqnarray}
$s_{\tilde q}=\sin{\theta_{\tilde q}}$ and $c_{\tilde
q}=\cos{\theta_{\tilde q}}$ with $\theta_{\tilde q}$ being the
$\tilde q_L-\tilde q_R$ mixing angle. The coefficients $g_{iL}^q$
and $g_{iR}^q$ in the mass eigenstate basis are
\begin{eqnarray}
&& g_{1L}^{q} = c_{\tilde q}\, g_{LL}^{q}
             + s_{\tilde q}\, g_{LR}^{q},\quad \,
   g_{1R}^{q} = c_{\tilde q}\, g_{RL}^{q}
             + s_{\tilde q}\, g_{RR}^{q}, \\[2mm]
&& g_{2L}^{q} =- s_{\tilde q}\, g_{LL}^{q} + c_{\tilde
q}\,g_{LR}^{q},\quad
   g_{2R}^{q} = - s_{\tilde q}\, g_{RL}^{q}+
   c_{\tilde q}\, g_{RR}^{q}\,.
\end{eqnarray}
The $g_{LL}^q, g_{LR}^q, g_{RR}^q$ and $g_{RL}^q$ coefficients for
the up-type quarks $u$, $c$ and $t$ are
\begin{eqnarray}
&& g_{LL}^{u}= -\frac{m_u}{2M_Ws_\beta t_W} N_{14}^* , \quad
g_{LR}^{u}=\frac{2}{3}\left( c_W N_{11}^{\prime *}-s_W
N_{12}^{\prime *} \right),  \\[1mm] && g_{RR}^{u} =
-\frac{m_u}{2M_Ws_\beta t_W} N_{14},  \quad
g_{RL}^{u}=-\frac{2}{3} c_W\,
N_{11}^\prime-(\frac{1}{2}+\frac{2}{3}s_W^2)\,
\frac{N_{12}^\prime}{s_W}~~~~
\end{eqnarray}
and for the down-type quarks $d$, $s$ and $b$ are
\begin{eqnarray}
&& g_{LL}^{d}= -\frac{m_d}{2M_Wc_\beta t_W} N_{13}^* ,\quad
g_{LR}^{d}=-\frac{1}{3}\left( c_W N_{11}^{\prime *}-s_W
N_{12}^{\prime *} \right),  \\[1mm] && g_{RR}^{d} =
-\frac{m_d}{2M_Wc_\beta t_W} N_{13},  \quad g_{RL}^{d}=\frac{1}{3}
c_W\, N_{11}^\prime+\left(\frac{1}{2}-\frac{1}{3}s_W^2\right)\,
\frac{N_{12}^\prime}{s_W},~~~~
\end{eqnarray}
where $s_W=\sin{\theta_{W}},\>c_W=\cos{\theta_W}$,
$s_\beta=\sin\beta,\>c_\beta=\cos\beta$, $m_{u[d]}$ are the masses
of the up-type [down-type] quarks and $N$ is the matrix which
diagonalizes the neutralino mass matrix. Finally, $N_{11}^\prime$
and $N_{12}^\prime$ are defined as follows \cite{hg}:
\begin{eqnarray}
\llgm{N_{11}^{\prime}\cr N_{12}^{\prime}}\rrgm =\llgm{c_W & s_W
\cr -s_W & c_W }\rrgm \llgm{N_{11} \cr N_{12}}\rrgm.
\label{eq:rotationN}
\end{eqnarray}

\subsubsection{\bf Z-boson{\boldmath $-$}Quark{\boldmath $-$}Quark
Vertex {\rm (Ref.  \cite{hk}, Fig. 71)}}

The Lagrangian which describes the Z-boson$-$quark$-$quark
interaction is
\begin{eqnarray}
  {\cal L}_{Z q q} &=&  g_Z\, \bar q \,\gamma^\mu
   \Big( R_q\,P_R+L_q\,P_L\Big)q Z_\mu,
\end{eqnarray}
where $R_q$ and $L_q$ for the up-type [down-type] quarks read
\begin{eqnarray}
L_{u[d]}  = -[+]~\left(1-[+]~2\,\frac{2~[-1]}{3} s^2_W
\right)~~\mbox{and}~~ R_{u[d]}  =-2\,\frac{2~[-1]}{3} s^2_W,
\end{eqnarray}
and $g_Z=g/2 c_W$.

%%%%%%%%%

\subsubsection{Z-boson{\boldmath $-$}Neutralino{\boldmath $-$}Neutralino Vertex
{\rm (Ref.  \cite{hk}, Fig. 75)}}

The Lagrangian which describes the
Z-boson$-$neutralino$-$neutralino interaction is
\begin{eqnarray}
   {\cal L}_{Z\tilde\chi\tilde\chi}& =& g_{
   Z\tilde\chi\tilde\chi}\,
\bar{\tilde\chi} \gamma^\mu\gamma_5\tilde\chi \,Z_\mu \, ,
~~\mbox{where}~~ g_{ Z\tilde\chi\tilde\chi} = {g_Z\over 2}\,
\left(|N_{13}|^2 - |N_{14}|^2 \right).~~~~~~
\end{eqnarray}

%%%%%%%%%%%%%%%%%%%%%%%%%%%%%%%%%%%

\subsubsection{Higgs{\boldmath $-$}Quark{\boldmath $-$}Quark Vertex
{\rm (Ref.  \cite{hg}, Fig. 7)}}

The Lagrangian which describes the Higgs$-$quark$-$quark
interaction is
\begin{eqnarray}
{\cal L}_{h,H,A~ qq} &=& \bar q\, \Big( g_{hqq}\, h + g_{Hqq}\, H
+i g_{Aqq}\, \gamma_{5}\, A \Big) \, q,
\end{eqnarray}
where the $g_{hqq}, g_{Hqq}$ and $g_{Aqq}$ coefficients for the
up- and down-type quarks are
\begin{eqnarray}
g_{huu}=-\frac{g\,m_u c_\alpha}{2\,M_W s_\beta}, &&
g_{hdd}=\frac{g\,m_d s_\alpha}{2\,M_W c_\beta},\\
g_{Huu}=-\frac{g\,m_u s_\alpha}{2\,M_W s_\beta}, &&
g_{Hdd}=-\frac{g\,m_d c_\alpha}{2\,M_W c_\beta},\\
%\mbox{and}~~
g_{Auu}=-\frac{g\,m_u }{2\,M_W } \tan^{-1}\beta, &&
g_{Add}=-\frac{g\,m_d }{2\,M_W }\tan\beta\,,~~~~~~
\end{eqnarray}
where $c_\alpha=\cos\alpha$ and $s_\alpha=\sin\alpha$ with
$\alpha$ being the Higgs mixing angle \cite{hg}.

%%%%%%%%%%%%%%%%%%%%%%%%%%%%%%%%%%%

\subsubsection{Higgs{\boldmath $-$}Neutralino{\boldmath $-$}Neutralino Vertex {\rm (Ref.
\cite{hg}, Fig. 21)}}

The Lagrangian which describes the Higgs$-$neutralino$-$neutralino
interaction is
\begin{eqnarray}
   && {\cal L}_{h,H,A~ \tilde\chi\tilde\chi}=
{1\over 2}\, \bar{\tilde\chi} \Big(g_{h\tilde\chi\tilde\chi}\, h+
g_{H\tilde\chi\tilde\chi}\, H +i g_{A\tilde\chi\tilde\chi}\,
\gamma_5\,A \Big)\tilde\chi\,,
\end{eqnarray}
where
\begin{eqnarray}
g_{h[H]\tilde\chi\tilde\chi}=g\Big(
s_\alpha\,[-c_\alpha]\,Q_{11}+c_\alpha\, [s_\alpha]\,S_{11}\Big)~~
\mbox{and}~~g_{A\tilde\chi\tilde\chi}=g\Big(
s_\beta\,Q_{11}-c_\beta\,S_{11}\Big)\hfill
\end{eqnarray}
with
\begin{eqnarray}
Q_{11}=N_{13}\left(N_{12}-t_W N_{11}\right)~~ \mbox{and}~~
S_{11}=N_{14}\left(N_{12}-t_W N_{11}\right).
\end{eqnarray}

%--------------------------------
\subsection{Neutralino{\boldmath $-$}Quark Effective Lagrangian}\label{eff}
%--------------------------------involving

From the expressions above, it is possible \cite{early, drees1,
sgmCP}, after the appropriate Fierz rearrangement, to derive the
coefficients $\alpha^{\rm SI}_q$ and $\alpha^{\rm SD}_q$ contained
in the effective four-fermion Lagrangian for the
neutralino$-$quark elastic scattering:
%%%%%%%%%%%%%%%%%%%%%%%%%%%%%%%%%%%%%%%%%%%%%%%%%%%%%%%%%%%%%%%%%%%%
\begin{figure}[!t]
\centerline{\epsfig{file=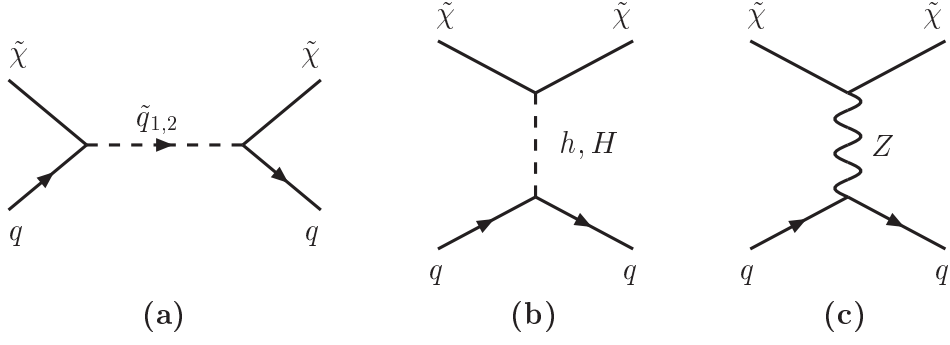,height=4.5cm}} \caption{\sl\ftn
The five Feynman diagrams contributing to the neutralino$-$quark
elastic scattering process $\tilde\chi\, q \rightarrow\tilde\chi\,
q$: (a) the two diagrams with squark $\tilde{q}_{1,2}$ exchange,
(b) the two diagrams with spin-0 neutral-Higgs-boson $(h, H)$
exchange, and (c) the diagram with spin-1 $Z$-boson exchange.}
\label{xqxq}
\end{figure}
%%%%%%%%%%%%%%%%%%%%%%%%%%%%%%%%%%%%%%%%%
%
\begin{eqnarray}\label{lnqeff}
{\cal L}_{\rm eff}=\alpha^{\rm
SI}_q\,\left(\bar{\tilde\chi}\tilde\chi\right)\,
(\bar{q}q)+\alpha^{\rm SD}_q\,\left(\bar{\tilde\chi}
\gamma^\mu\gamma_5 \tilde\chi\right)\,\left(\bar{q} \gamma_\mu
\gamma_5 q\right)+\cdots,
\end{eqnarray}
where the ellipsis represents (i) terms (such as
$\bar{\tilde\chi}\gamma_5\tilde\chi\,\bar{q} \gamma_5 q$) which
generate velocity-dependent and, thus, negligible \cite{sgmCP}
contributions to the elastic neutralino$-$quark cross section,
(ii) loop corrections of order $m_{\tilde q_{1,2}}^{-4}$ which
were previously considered in the literature \cite{drees1} and can
be safely neglected for $m_{\tilde q_{1,2}}\gg m_{\tilde\chi}$
\cite{kov,efo} (see below). The five tree-level Feynman diagrams
contributing to the neutralino$-$quark scattering process are
shown in Fig.~\ref{xqxq}. The $s$-channel squark-exchange diagrams
contribute to both $\alpha^{\rm SI}_q$ and $\alpha^{\rm SD}_q$,
while the $t$-channel Higgs-exchange diagrams only to $\alpha^{\rm
SI}_q$. The $t$-channel $Z$-exchange diagram contributes only to
$\alpha^{\rm SD}_q$. These coefficients, in the non-relativistic
limit, read
\begin{eqnarray}
\label{sind} \alpha^{\rm SI}_q = -g^2t_{{W}}^2\sum_{i=1,2}
\frac{g^q_{i L}\,g^q_{i R}}{m_{\tilde q_i}^2 - (m_{\tilde\chi}+m_q
)^2} + \frac{g_{{h\tilde\chi\tilde\chi}}\,g_{{hqq}}}{2m^2_h}+
\frac{g_{{H\tilde\chi\tilde\chi}}\,g_{{Hqq}}}{2m^2_H},
\\ \label{sda}\alpha^{\rm SD}_q = -\frac{g^2t_{_{W}}^2}{2} \sum_{i=1,2}
\frac{\left|\, g^q_{i L}\, \right|^2 + \left|\, g^q_{i R}\right|^2
} { m_{\tilde q_i}^2 - (m_{\tilde\chi} +m_q )^2
}-\frac{g_{{Z}}^2}{2M_Z^2}g_{{Z\tilde\chi\tilde\chi}}
(R_q-L_q).~~~
\end{eqnarray}

Finally, let us note that, besides the neutralino$-$quark
interaction in Eq. (\ref{lnqeff}), there are also
neutralino$-$gluon interactions with contributions arising from
one-loop heavy quark and Higgs diagrams and the so-called twist-2
operators \cite{drees1, jungman}. However, these corrections are
negligible for the values of SUSY parameters encountered in our
$\mu>0$ model (see Sec. \ref{phenoaa}).

%-------------------------------------------------------------------
\subsection{Neutralino{\boldmath $-$}Nucleus Spin Independent Cross
Section}\label{si}
%-------------------------------------------------------------------

The SI part of $\sigma^0_{\tilde\chi {\rm N}}$ can be
parameterized as
\begin{eqnarray}\label{sSI}
\sigma_{\tilde\chi{\rm N}}^{\rm
SI}=\frac{4}{\pi}\,\mu^2_{\tilde\chi {\rm N}}\Big(\,Z_{\rm
N}f_p+(A_{\rm N}-Z_{\rm
N})\,f_n\,\Big)^2,~~\mbox{where}~~\mu_{\tilde\chi {\rm
N}}=\frac{m_{\tilde{\chi}}m_{\rm N}}{m_{\tilde\chi}+m_{\rm N}}
\end{eqnarray}
is the neutralino$-$nucleus reduced mass and $Z_{\rm N}$ and
$A_{\rm N}$ denote the atomic number and the mass number of the
nucleus respectively. In the limit of $m_{\tilde\chi} \ll
m_{\tilde q_{1,2}}$ (which is, in general, true unless $\tilde
q_2$ is the next-to-LSP coinciding with the lightest stop
\cite{drees, boem} or sbottom \cite{su5b} quark) and $m_{q} \ll
m_{\tilde q_{1,2}}$, the effective coupling $f_{p[n]}$ of the LSP
to proton [neutron] is given by
\begin{eqnarray}\label{fpn}
f_{p[n]}=\sum_{q=u,d,s} \frac{m_{p[n]}}{m_q} f_{_{{\rm
T}_q}}^{p[n]}\, \alpha^{\rm SI}_q +\frac{2}{27}\,f_{_{{\rm
T}_G}}^{p[n]}\sum_{q=c,b,t} \frac{m_{p[n]}}{m_q}\, \alpha^{\rm
SI}_q
\end{eqnarray}
to lowest order in  $m^{-1}_{\tilde q_{1,2}}$, where $m_{p[n]}$ is
the proton [neutron] mass and the parameters $f_{_{{\rm
T}_q}}^{p}$ and $f_{_{{\rm T}_q}}^{n}$ are taken to be \cite{efo}
\begin{eqnarray}
&& f_{_{{\rm T}_u}}^p=0.02\pm0.004,~
   f_{_{{\rm T}_d}}^p=0.026\pm0.005,~
   f_{_{{\rm T}_s}}^p=0.118\pm0.062,
   \label{rgis} \\
&& f_{_{{\rm T}_u}}^n=0.014\pm0.003,~
   f_{_{{\rm T}_d}}^n=0.036\pm0.008,~
   f_{_{{\rm T}_s}}^n=0.118\pm0.062.~~~~~~~~~~
   \label{fTn}
\end{eqnarray}
with $f_{_{{\rm T}_G}}^{p[n]}=1-\sum_{q=u,d,s}f_{_{{\rm
T}_q}}^{p[n]}$. In our numerical analysis, we use the running
masses of $b$ and $t$ quark at the scale $m_{\tilde\chi}$.

%-------------------------------------------------------------------
\subsection{Neutralino{\boldmath $-$}Nucleus Spin Dependent Cross
Section}\label{sd}
%-------------------------------------------------------------------

The SD part of $\sigma^0_{\tilde\chi {\rm N}}$ can be
parameterized as follows:
\begin{eqnarray}\label{sSD}
\sigma_{\tilde\chi {\rm N}}^{\rm SD}=\frac{16}{\pi}\,
\mu^2_{\tilde\chi {\rm N}}\,\Lambda^2\, J_{\rm N}(J_{\rm
N}+1),~~\mbox{where}~~\Lambda =\frac{1}{J_{\rm N}}\Big(\,\lambda_p
\,\langle S_p \rangle +\lambda_n \, \langle S_n \rangle\Big)
\end{eqnarray}
and $J_{\rm N}$ is the total angular momentum of the nucleus
($9/2$ for ${^{73}}{\rm Ge}$). Also, $\langle S_{p[n]} \rangle$ is
the expectation value of the spin content of the proton [neutron]
group in the nucleus with explicit value, for a ${^{73}}{\rm Ge}$
target, $\langle S_{p}\rangle_{_{^{73}{\rm Ge}}}=0.011~[\langle
S_{n}\rangle_{_{^{73}{\rm Ge}}}=0.491]$ in the shell model.
Finally, the coefficient $\lambda_{p[n]}$ is parameterized in
terms of the quark spin content of the proton [neutron]
$\Delta_q^{p[n]}$ and the effective axial-vector couplings
$\alpha^{\rm SD}_q~(q=u,\, d,\, s)$ as
\begin{eqnarray}\label{Dqn}
\lambda_{p[n]} = \sum_{q=u,d,s} \alpha^{\rm
SD}_q\,\Delta_q^{p[n]}\,,~~\mbox{where}~~\Delta_{u}^{n}=\Delta_{d}^{p},~
\Delta_{d}^{n}=\Delta_{u}^{p}~~\mbox{and}~~\Delta_{s}^{n}=\Delta_{s}^{p}
\end{eqnarray}
with the factors $\Delta_q^{p}$ taken to be \cite{efo}
\begin{eqnarray}
\label{Dqp} \Delta_{u}^{p}=+0.78\pm0.02,~
\Delta_{d}^{p}=-0.48\pm0.02~~\mbox{and}~~
\Delta_{s}^{p}=-0.15\pm0.02.
\end{eqnarray}
Needless to say that the formalism of this section remains valid
even in the case of light stop \cite{drees, boem} or sbottom
\cite{su5b} quarks.

%%%%%%%%%%%%%%%%%%%%%%%%%%%%%%%%%%%%%%%%%%%%%%%%%%%%%%%%%%%%

\def\ijmp#1#2#3{{Int. Jour. Mod. Phys.}
{\bf #1},~#3~(#2)}
\def\plb#1#2#3{{Phys. Lett. B }{\bf #1},~#3~(#2)}
\def\zpc#1#2#3{{Z. Phys. C }{\bf #1},~#3~(#2)}
\def\prl#1#2#3{{Phys. Rev. Lett.}
{\bf #1},~#3~(#2)}
\def\rmp#1#2#3{{Rev. Mod. Phys.}
{\bf #1},~#3~(#2)}
\def\prep#1#2#3{{Phys. Rep. }{\bf #1},~#3~(#2)}
\def\prd#1#2#3{{Phys. Rev. D }{\bf #1},~#3~(#2)}
\def\npb#1#2#3{{Nucl. Phys. }{\bf B#1},~#3~(#2)}
\def\npps#1#2#3{{Nucl. Phys. B (Proc. Suppl.)}
{\bf #1},~#3~(#2)}
\def\mpl#1#2#3{{Mod. Phys. Lett.}
{\bf #1},~#3~(#2)}
\def\arnps#1#2#3{{Annu. Rev. Nucl. Part. Sci.}
{\bf #1},~#3~(#2)}
\def\sjnp#1#2#3{{Sov. J. Nucl. Phys.}
{\bf #1},~#3~(#2)}
\def\jetp#1#2#3{{JETP Lett. }{\bf #1},~#3~(#2)}
\def\app#1#2#3{{Acta Phys. Polon.}
{\bf #1},~#3~(#2)}
\def\rnc#1#2#3{{Riv. Nuovo Cim.}
{\bf #1},~#3~(#2)}
\def\ap#1#2#3{{Ann. Phys. }{\bf #1},~#3~(#2)}
\def\ptp#1#2#3{{Prog. Theor. Phys.}
{\bf #1},~#3~(#2)}
\def\apjl#1#2#3{{Astrophys. J. Lett.}
{\bf #1},~#3~(#2)}
\def\n#1#2#3{{Nature }{\bf #1},~#3~(#2)}
\def\apj#1#2#3{{Astrophys. J.}
{\bf #1},~#3~(#2)}
\def\anj#1#2#3{{Astron. J. }{\bf #1},~#3~(#2)}
\def\mnras#1#2#3{{MNRAS }{\bf #1},~#3~(#2)}
\def\grg#1#2#3{{Gen. Rel. Grav.}
{\bf #1},~#3~(#2)}
\def\s#1#2#3{{Science }{\bf #1},~#3~(#2)}
\def\baas#1#2#3{{Bull. Am. Astron. Soc.}
{\bf #1},~#3~(#2)}
\def\ibid#1#2#3{{\it ibid. }{\bf #1},~#3~(#2)}
\def\cpc#1#2#3{{Comput. Phys. Commun.}
{\bf #1},~#3~(#2)}
\def\astp#1#2#3{{Astropart. Phys.}
{\bf #1},~#3~(#2)}
\def\epjc#1#2#3{{Eur. Phys. J. C}
{\bf #1},~#3~(#2)}
\def\nima#1#2#3{{Nucl. Instrum. Meth. A}
{\bf #1},~#3~(#2)}
\def\jhep#1#2#3{{J. High Energy Phys.}
{\bf #1},~#3~(#2)}
\def\jcap#1#2#3{{J. Cosmol. Astropart. Phys.}
{\bf #1},~#3~(#2)}

\end{document}